\documentclass[12pt]{article}

\usepackage{scicite}

\usepackage{times}

\usepackage{siunitx}
\usepackage{graphicx}
\usepackage{lscape}
\usepackage{booktabs}
\usepackage{multirow}
\usepackage[version=4]{mhchem}
\usepackage{amsmath,amssymb}

\newenvironment{sciabstract}{%
\begin{quote} \bf}
{\end{quote}}

\title{Discovery of interstellar 1-cyanopyrene: a four-ring polycyclic aromatic hydrocarbon in TMC-1}

\author
{Gabi Wenzel$^{1\ast}$, 
Ilsa R. Cooke$^{2\ast}$, 
P. Bryan Changala$^{3}$,\\
Edwin A. Bergin$^{4}$,
Shuo Zhang$^{1}$,
Andrew M. Burkhardt$^{5}$,\\
Alex N. Byrne$^{1}$,
Steven B. Charnley$^{6}$,
Martin A. Cordiner$^{6}$,\\
Miya Duffy$^{1}$,
Zachary T. P. Fried$^{1}$,
Harshal Gupta$^{3,7}$,\\
Martin S. Holdren$^{1}$,
Andrew Lipnicky$^{8}$,
Ryan A. Loomis$^{8}$,\\
Hannah Toru Shay$^{1}$,
Christopher N. Shingledecker$^{9}$,
Mark A. Siebert$^{10}$, \\
D. Archie Stewart$^{1}$,
Reace H. J. Willis$^{2}$,
Ci Xue$^{1}$,\\
Anthony J. Remijan$^{8}$,
Alison E. Wendlandt$^{1}$,\\
Michael C. McCarthy$^{3}$,
Brett A. McGuire$^{1,8\ast}$,\\
\footnotesize
$^{1}$Department of Chemistry, Massachusetts Institute of Technology, Cambridge, MA 02139, USA.\\
\footnotesize
$^{2}$Department of Chemistry, University of British Columbia, Vancouver, BC V6T 1Z1, Canada.\\
\footnotesize
$^{3}$Center for Astrophysics $\mid$ Harvard~\&~Smithsonian, Cambridge, MA 02138, USA.\\
\footnotesize
$^{4}$Department of Astronomy, University of Michigan, Ann Arbor, MI 48109, USA.\\
\footnotesize
$^{5}$Department of Earth, Environment, and Physics, Worcester State University, Worcester, MA 01602, USA.\\
\footnotesize
$^{6}$Astrochemistry Laboratory, NASA Goddard Space Flight Center,\\ 
\footnotesize
8800 Greenbelt Road, Greenbelt, MD 20771, USA.\\
\footnotesize
$^{7}$National Science Foundation, Alexandria, VA 22314, USA.\\
\footnotesize
$^{8}$National Radio Astronomy Observatory, Charlottesville, VA 22903, USA.\\
\footnotesize
$^{9}$Department of Chemistry, 
Virginia Military Institute, Lexington, VA
24450, USA.\\
\footnotesize
$^{10}$Department of Astronomy, University of Virginia, Charlottesville, VA 22904, USA.
\\
\normalsize{$^\ast$Corresponding authors. E-mail: gwenzel@mit.edu, icooke@chem.ubc.ca, brettmc@mit.edu.}
}

\date{}

\begin{document} 


\baselineskip24pt


\maketitle 

\begin{sciabstract}
Polycyclic aromatic hydrocarbons (PAHs) are expected to be the most abundant class of organic molecules in space. Their interstellar lifecycle is not well understood, and progress is hampered by difficulties detecting individual PAH molecules. Here, we present the discovery of \ce{CN}-functionalized pyrene, a 4-ring PAH, in the dense cloud TMC-1 using the 100-m Green Bank Telescope. We derive an abundance of 1-cyanopyrene of $\mathbf{{\sim}1.52 \times 10^{12}\,cm^{-2}}$, and from this estimate that the un-substituted pyrene accounts for up to $\mathbf{0.03-0.3\,\%}$ of the carbon budget in the dense interstellar medium which trace the birth sites of stars and planets. The presence of pyrene in this cold ($\sim$10\,K) molecular cloud agrees with its recent measurement in asteroid Ryugu where isotopic clumping suggest a cold, interstellar origin. The direct link to the birth site of our solar system is strengthened when we consider the solid state pyrene content in the pre-stellar materials compared to comets, which represent the most pristine material in the solar system. We estimate that solid state pyrene can account for 1\,\% of the carbon within comets carried by this one single organic molecule. The abundance indicates pyrene is an ``island of stability" in interstellar PAH chemistry and suggests a potential cold molecular cloud origin of the carbon carried by PAHs that is supplied to forming planetary systems, including habitable worlds such as our own.
\end{sciabstract}

Unidentified infrared (UIR) bands have been observed for decades, dominating the infrared spectra of many dust-rich astronomical objects in our galaxy and beyond \cite{tielens2008}. It is widely accepted that the carriers of this emission are a particular class of exceptionally stable molecules, polycyclic aromatic hydrocarbons (PAHs) \cite{allamandola1985,leger1984,allamandola1989}. PAHs are thought to be present at many stages of the stellar life cycle and their stability against ultraviolet (UV) radiation makes them promising candidates to survive even under harsh interstellar conditions \cite{chabot2020}. The UIR bands that have been assigned to vibrational modes of PAHs are found predominantly in regions illuminated by hot stars or the interstellar radiation field.  Recent \emph{James Webb Space Telescope} (JWST) observations of these photon-dominated regions (PDRs) \cite{peeters2011} have studied the UIRs at high sensitivity and spectral resolution \cite{chown2024}. While no individual PAHs have been identified via the UIR features, these observations show that these molecules are among the most ubiquitous organic compounds in the universe.  The intensity of their UIR bands indicates that as much as ${\sim}10-25\,\%$ of all carbon in the Galactic interstellar medium may be incorporated into PAHs \cite{dwek1997,habart2004a,tielens2008,chabot2020}. PAHs have also been suggested to be carriers of the diffuse interstellar bands (DIBs)\cite{vanderzwet1985,leger1985}, ubiquitous absorption features in astronomical spectra extending from the UV into near-IR wavelengths, whose chemical origins remain a mystery --- with the notable exception of \ce{C60+} \cite{campbell2015,cordiner2019}. The pyrene cation, \ce{C16H10+}, has been proposed as a possible DIB carrier due to a strong optical absorption feature (measured in rare gas matrices) that lies close to the DIB at 4430\,\r{A}\cite{salama1992}; however, unambiguous identification of \ce{C16H10+}, or any other PAH ion as carriers of the DIBs remains elusive.

In the 1960s, PAHs were identified in carbonaceous chondrites, remnants of our own presolar nebula \cite{studier1965}. More recent findings -- including their presence in samples from comets \cite{clemett2010} and asteroids \cite{zeichner2023} -- suggest that the PAHs found in these reservoirs can be traced even farther back, to an origin in the primordial dense cloud that predates our presolar nebula. The abundance ratio of stable carbon isotopes ($^{13}$C/$^{12}$C) is consistent with an extraterrestrial origin for these PAHs, which likely formed by kinetically controlled mass growth processes \cite{naraoka2000}.
Many models of PAH formation are derived from those developed in the research of combustion systems, primarily involving reactions with substantial activation energies that are thought to take place in the high-temperature ($>\!\!1000\,\mathrm{K}$) regions of circumstellar envelopes around asymptotic giant branch (AGB) stars \cite{zhao2018} which qualitatively resemble combustion conditions. Nevertheless, such models may not fully account for the abundance of PAHs in the interstellar medium (ISM). In particular, small PAHs ($\lesssim\!35$ C atoms) are expected to be destroyed by shock waves, galactic cosmic rays, and ultraviolet photons more rapidly than the estimated injection rate of new PAHs from circumstellar envelopes into the interstellar medium \cite{micelotta2011}. Despite this, 1- and 2-cyanonaphthalene, cyano (-CN) derivatives of the smallest 2-ring PAH naphthalene, as well as indene and 2-cyanoindene, have been detected in the Taurus molecular cloud in the in the cold, starless molecular cloud TMC-1\cite{mcguire2021,burkhardt2021,cernicharo2021,sita2022}.

Similar to the fullerenes \ce{C60} and \ce{C70} that have been detected toward planetary and reflection nebulae \cite{cami2010,sellgren2010}, pyrene is considered to represent an ``island of stability'' --- an especially thermodynamically and kinetically stable molecule whose formation is practically irreversible when produced via combustion \cite{frenklach2024}.  The high abundance of pyrene relative to other PAHs found in carbonaceous chondrites, such as Murchison and Orgeuil \cite{sabbah2017,lecasble2022,aponte2023} and, recently, the asteroid Ryugu \cite{zeichner2023}, support this stability argument. Although high-temperature routes to pyrene have been proposed \cite{zhao2018}, these formation mechanisms appear inconsistent with recent carbon-isotope analysis of samples from Ryugu. Instead, $^{13}$C-isotope clumping suggests that the pyrene isolated from Ryugu formed in cold interstellar environments. While some barrierless low-temperature mechanisms to form these small PAHs have been proposed, conventional chemical models have so far been unable to reproduce their observed abundances \cite{byrne2023}. The results from Ryugu, combined with the large abundance of pyrene observed in solar system materials, and the recent detections of the smallest PAHs in TMC-1, motivated our search for interstellar cyanopyrene. 

\begin{figure}
    \centering
    \includegraphics[width=0.3\linewidth]{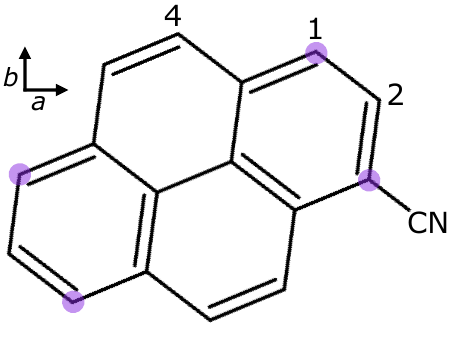}
    \caption{\textbf{Molecular structure of 1-cyanopyrene.} \ce{C17H9N} or \ce{1-CN}-\ce{C16H9}, a derivative of the pericondensed, compact PAH pyrene, \ce{C16H10}. Substituting a H atom on the 1-position of pyrene for a nitrile-group (\ce{-CN}) introduces a strong permanent dipole moment along the $a$- and $b$-components of its principal axis system of $\mu_a = 4.8\,\mathrm{D}$ and $\mu_b = 2.3\,\mathrm{D}$, respectively. The colored dots denote equivalent substitution sites of pyrene to form the same 1-cyanopyrene isomer.}
    \label{fig:molecule_structure}
\end{figure}

\begin{figure}[!htb]
    \centering
    \includegraphics[width=\textwidth]{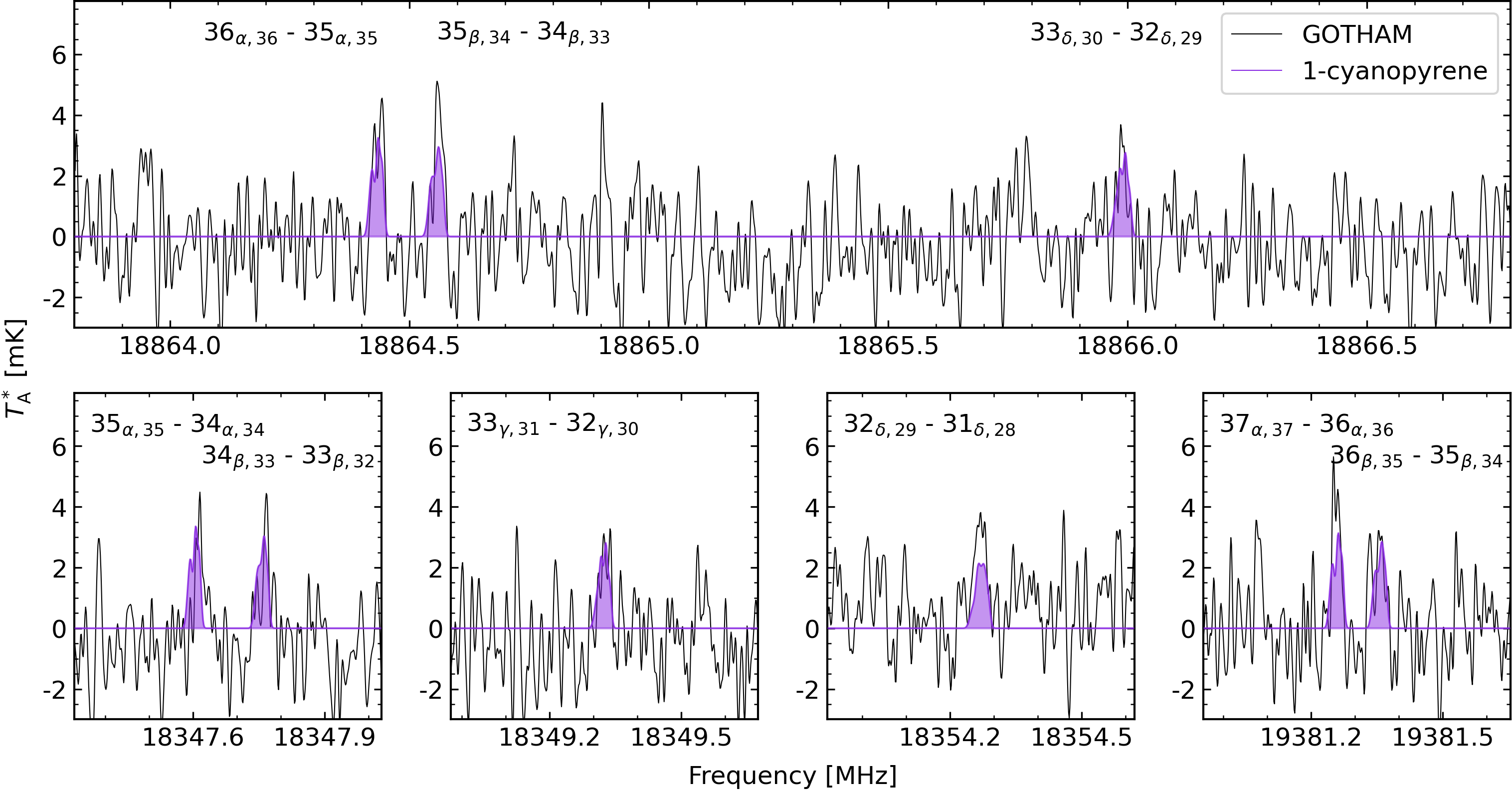}
    \caption{\textbf{GOTHAM spectra showcasing several 1-cyanopyrene lines detected in TMC-1 observations.} The original GOTHAM observational data for which one channel corresponds to $1.4\,\mathrm{kHz}$ were smoothed with a Hanning window to a resolution of $14\,\mathrm{kHz}$, depicted in black. The spectrum of 1-cyanopyrene is overplotted in violet using source-dependent molecular parameters as reported in Table \ref{tab:MCMCresults}. The quantum numbers of each transition, ignoring $^{14}$N nuclear electric quadrupole splitting, are reported. Each line contains multiple closely spaced $K$-components of each transition which are denoted $\alpha, \beta, \gamma, \delta = \{0,1\}, \{1,2\}, \{2,3\}, \{3,4\}$.}
    \label{fig:indilines}
\end{figure}

\subsection*{Discovery of 1-cyanopyrene in the GOTHAM survey of TMC-1}

We analyzed a high-sensitivity, high-spectral resolution broadband line survey of TMC-1 with near-continuous coverage from approximately $8$ to $36\,\mathrm{GHz}$ collected with the 100-m Robert C. Byrd Green Bank Telescope (GBT) as part of the GBT Observations of TMC-1: Hunting Aromatic Molecules (GOTHAM) project \cite{mcguire2020}. Multiple new interstellar species have been discovered with the GOTHAM data set, including the unambiguous detection of the 1- and 2-cyanonaphthalene isomers \cite{mcguire2021}, indene\cite{burkhardt2021}, and 2-cyanoindene\cite{sita2022}. A permanent dipole is a prerequisite for detection with rotational spectroscopy, a technique that has been used to discover more than $90\,\%$ of interstellar species \cite{mcguire2022}. PAHs, due to their symmetry and delocalized $\pi$-electrons across the aromatic rings, commonly possess a weak or non-existent dipole moment, meaning that they cannot be readily observed using radio astronomy and must be indirectly detected through chemically related molecules. Laboratory work has shown that \ce{CN}-functionalized aromatics can be used as efficient proxies when searching for pure hydrocarbon species that do not possess permanent dipole moments \cite{balucani2000,cooke2020}. 

Due to its enhanced stability \cite{leger1987,naraoka2023}, pyrene, the smallest pericondensed (or compact) PAH, in which all rings are connected to at least two others, has been the target of many experimental and theoretical studies \cite{jusko2018a,marin2020,campisi2020,jlee2021,wenzel2022}. Substituting one H atom bonded to a sp$^2$-hybridized edge carbon for a nitrile (\ce{-CN}) group will introduce a strong permanent dipole moment to the otherwise rotationally invisible pyrene. Three unique cyanopyrene isomers can be produced (see Fig. \ref{fig:molecule_structure}), but to our knowledge, no prior laboratory rotational spectroscopy existed for these species. Quantum chemical calculations of the lowest-energy isomer, 1-cyanopyrene \cite{garkusha2011}, predicted strong transitions in the 8--22\,GHz frequency range \cite{ye2022}, making this molecule an excellent candidate for a search in the GOTHAM data set \cite{sita2022,cooke2023,Science:Materials}.

We measured the high-resolution laboratory rotational spectra for 1-cyanopyrene to enable a search for it in our astronomical observations. 1-cyanopyrene was synthesized from pyrene in a gallium-catalyzed cyanation reaction \cite{okamoto2012} as a white solid (see figs. \ref{fig:synthesis} and \ref{fig:NMR}). It was then evaporated in a laser ablation-supersonic expansion source, and its spectrum was collected with a cavity-enhanced Fourier transform microwave spectrometer \cite{Science:Materials}. The spectroscopic assignments were guided by quantum chemical predictions of the rotational constants \cite{Science:Materials}. We observed 267 individual or partly blended rotational transitions over the 7--16\,GHz frequency range directly overlapping with our GOTHAM observations, which were readily fit to a standard rotational Hamiltonian \cite{Science:Materials}. 

TMC-1 is known to contain four distinct, partially overlapping velocity components \cite{loomis2021}. A Markov chain Monte Carlo (MCMC) analysis was used to determine the column density of 1-cyanopyrene in each component, with tight prior constraints on the velocities, temperature, and linewidth informed by the prior detections of PAHs \cite{mcguire2021} in this source (Table \ref{tab:MCMCresults}). From the MCMC analysis, we derive a total column density as the sum of the column densities in all four Doppler components of $N_T(\mathrm{Total}) = 1.52^{+0.18}_{-0.16}\times 10^{12}\,\mathrm{cm}^{-2}$ (corresponding to an abundance of $N(\ce{C17H9N})/N(\ce{H2}) \approx 1.5 \times 10^{-10}$ \cite{gratier2016}) and a single excitation temperature of $7.87\,\mathrm{K}$.  The rotational spectrum of 1-cyanopyrene simulated using the MCMC-determined parameters using the \texttt{molsim} package \cite{molsim} is depicted in violet in Fig. \ref{fig:indilines} and shows numerous individually detected transitions.  

\begin{figure}
    \centering
    \includegraphics[width=\textwidth]{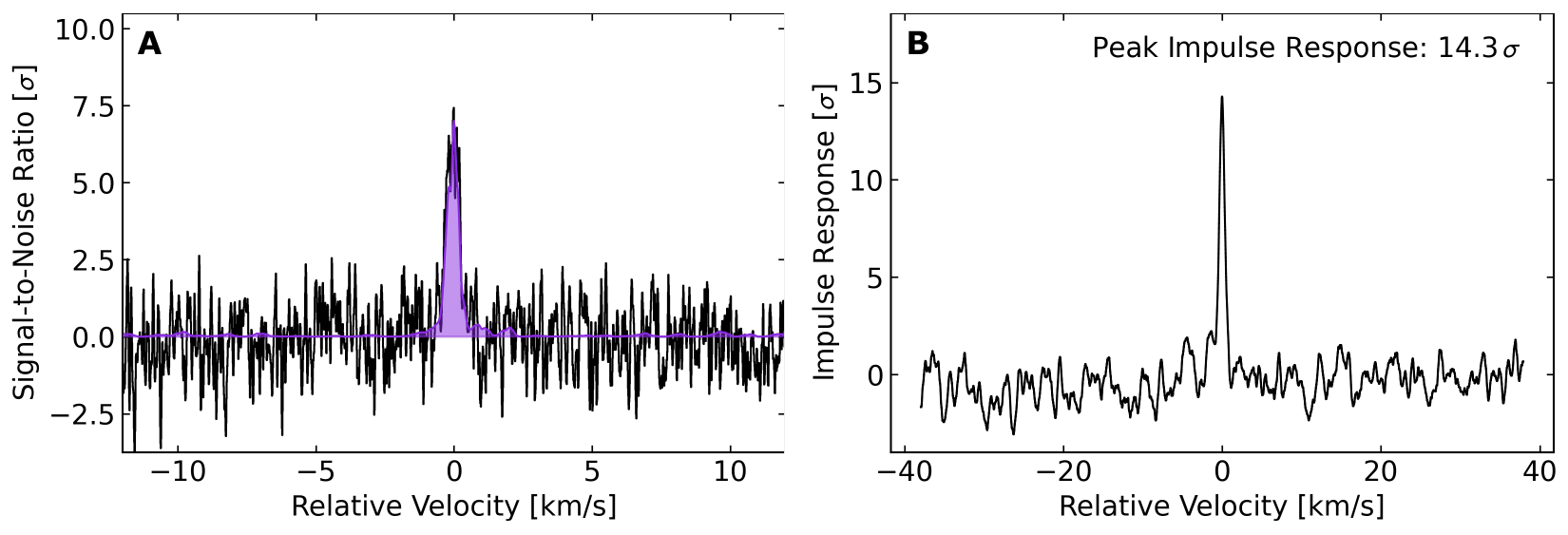}
    \caption{\textbf{Velocity-stacked spectra and matched filter response of 1-cyanopyrene.} The stacked GOTHAM observations (black) are overlaid with the 1-cyanopyrene stacked spectrum (violet) for the 150 brightest SNR 1-cyanopyrene lines (\textbf{A}). The corresponding impulse response for the matched filtering analysis is shown yielding a significance of $14.3\,\sigma$ for the 1-cyanopyrene detection (\textbf{B}).}
    \label{fig:stack+mf}
\end{figure}

In addition to the individual lines, we performed a velocity-stack and matched filtering analysis, which provides the total statistical evidence that the model of molecular emission from 1-cyanopyrene matches the observations. The process is described in detail elsewhere \cite{loomis2021,mcguire2021}. Briefly, small spectral windows, centered around each of the 150 brightest signal-to-noise ratio (SNR) 1-cyanopyrene lines in frequency space, are extracted and collapsed into one SNR weighted average line in velocity-space (Fig.~\ref{fig:stack+mf}A; \cite{Science:Materials}). Using the stacked, simulated spectrum as a matched filter generates an impulse response of $14.3\,\sigma$, representing the statistical significance of the detection \cite{loomis2021}. 

\subsection*{Implications for the carbon budget of TMC-1}

Since pyrene itself cannot be detected using radio astronomy, we estimate its abundance by considering the ratio of ``pure'' to CN-substituted aromatics, which we call the H:CN ratio. Based on the observed H:CN ratios for other related molecules in TMC-1, we derive an estimated abundance of pyrene of ${\sim}0.15-1.5\,\times 10^{-8}$ \cite{Science:Materials}. Given that pyrene consists of 16 carbon atoms, the abundance (i.e., relative to H) of carbon locked up in pyrene is between 0.02--0.2\,ppm. This amounts to $0.03$--$0.3\,\%$ of the carbon budget of TMC-1, assuming gas-phase $n_\mathrm{C}/n_\mathrm{H} = 9 \times 10^{-5}$ \cite{fuente2019}. Comparisons to other aromatic and cyclic species observed in TMC-1 are given in Table \ref{tab:budget}. Despite its size, cyanopyrene is more abundant than any of the CN-substituted aromatic/cyclic dienes, with the exception of benzonitrile, which is roughly equal in abundance. For comparison, the observed abundance of carbon in the cyanopolyynes (HC$_n$N, $n = 3-11$) in TMC-1 is $\sim$0.1\,ppm; thus pyrene locks up at least $20\,\%$ and up to double the amount of carbon as the cyanopolyynes. The abundance of pyrene relative to other molecules in TMC-1 suggests there may be a substantial reservoir of larger PAHs in the ISM.

The abundance of carbon locked up in PAHs has been estimated previously from the UIR bands from bright PDRs and the diffuse interstellar medium. A total carbon abundance of $14\,\mathrm{ppm}$ has been derived for PAHs with less than 100 carbon atoms \cite{tielens2008}. Assuming these PAHs consist of an average of 50 carbon atoms, the PAH abundance is $\sim3 \times 10^{-7}$. If the PAH population from dense molecular clouds is inherited during star formation, gas-phase pyrene may account for ${\sim}1\,\%$ of all PAHs in the presolar nebula. It is clear that there is a reservoir of carbon in TMC-1, in the form of PAH molecules that have yet to be observed. 

Since we are only detecting gas-phase (cyano)pyrene, we are blind to the amount condensed on dust grains. We estimate the abundance of solid-state pyrene locked up on dust grains is ${\sim}1\,\mathrm{ppm}$ (relative to H), based on its gas-phase abundance derived above and its expected depletion time under TMC-1 conditions \cite{Science:Materials}.  Comets are thought to represent the most pristine primordial material in our solar system and they may preserve the history of icy and refractory material locked up on dust grains. In contrast, rocky material, such as asteroid Ryugu, is depleted in carbon. To compare to the cometary carbon budget, we convert the solid-phase pyrene abundance from relative to H to relative to Si. Thus the carbon locked up in pyrene has an abundance of C/Si of 0.05 ppm, which amounts to ${\sim}1\,\%$ of the carbon content seen in comets \cite{bergin2015}. A high abundance of pyrene in solar system materials, inherited from interstellar clouds, is consistent with the findings from Ryugu and suggests a substantial inheritance between these stages.

\begin{figure}
    \centering
    \includegraphics[width=0.8\linewidth]{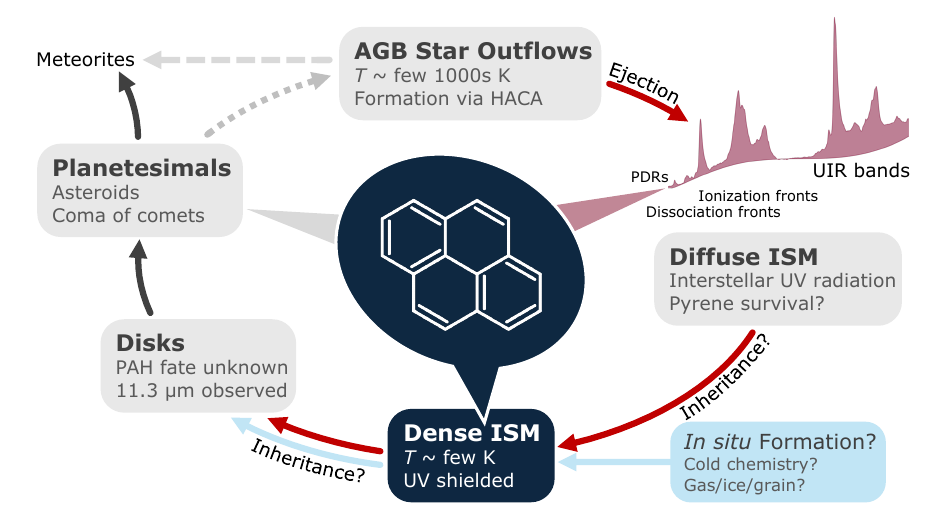}
     \caption{\textbf{Ubiquitous presence of PAHs closely connected to the star and cosmic dust lifecycles.} After PAHs are formed in circumstellar envelopes of AGB stars \cite{cernicharo2001,zhao2018}, they are ejected and seed the diffuse ISM. Here, PAHs are photoprocessed by the interaction with ultraviolet (UV) photons of the interstellar radiation field, which models have shown efficiently destroys PAHs with fewer than $50$ carbon atoms \cite{montillaud2013}. These astronomical objects are also sources of strong UIR emission \cite{chown2024}. While we detect 1-cyanopyrene in the cold, dense ISM as a proxy for the radio astronomically invisible pyrene, we cannot explain its high relative abundance through current inheritance or bottom-up formation mechanisms. This necessitates the exploration of possible \textit{in situ} formation routes at low temperatures. Recently, evidence for PAH emission at $11.3\,\mathrm{\mu m}$ has been found in a low-mass protoplanetary disk \cite{arulanantham2024}, from which finally planetesimals form. Identification of PAHs in carbonaceous chondrites, asteroids, and comets concludes the evidence of the ubiquitous presence of PAHs throughout the ISM.}
    \label{fig:cycle}
\end{figure}

\subsection*{Pyrene formation in TMC-1}

The detection of 1-cyanopyrene in TMC-1 adds to the growing number of PAHs that have been detected in cold molecular clouds. The questions remain: where did these PAHs form? Were they formed \textit{in situ} under the low-temperature conditions in TMC-1, or were they formed at higher temperatures in circumstellar envelopes and survived their journey through the diffuse ISM? Do these PAHs form through \textit{bottom-up} mechanisms involving molecular mass growth from small acyclic precursors or do they form by \textit{top-down} destruction of dust grains or the so-called grand-PAHs \cite{andrews2015}? 

High-temperature routes to PAHs, including pyrene, have been discussed in the literature, in part motivated by their connection to combustion systems on Earth. The most prominent of these is the Hydrogen Abstraction Acetylene (\ce{C2H2}) Addition (HACA) mechanism \cite{kaiser2021}. Acetylene has been shown to add to the 4-phenanthrenyl radical (\ce{C14H9}) to produce pyrene under conditions prevalent in carbon-rich circumstellar environments \cite{zhao2018}. Since this reaction has a sizable entrance barrier (${\sim}20\,\mathrm{kJ\,mol^{-1}}$), it is not expected to operate in TMC-1 or other low-temperature environments. Instead, an alternative barrierless mechanism involving hydrogen abstraction vinyl acetylene addition (HAVA) mechanisms has been proposed as a viable route to PAHs in dense clouds \cite{kaiser2021}. HAVA has been shown to be a viable route to naphthalene starting with phenyl \cite{parker2012,zhao2018a}; however, this reaction alone cannot reproduce or explain the observed abundance of the cyanonaphthalenes in astrochemical models of TMC-1 \cite{mcguire2021}. The HAVA route is also not a feasible route to pyrene since vinylacetylene will not add to the bay region (i.e., at carbons 4- and 5-) of phenanthrenyl.

If pyrene is formed in circumstellar envelopes, it must survive its journey through the diffuse interstellar medium as schematically depicted in Fig.\ref{fig:cycle}. Recent laboratory work has shown that fast radiative cooling may stabilize small cationic PAHs more efficiently than previously thought, due to the recurrent fluorescence mechanism involving the emission of optical photons from thermally populated electronically excited states \cite{stockett2023}. However, the inheritance of pyrene from circumstellar envelopes would be inconsistent with the $^{13}$C-clumping observed in pyrene from Ryugu, so alternative formation mechanisms need to be explored.

\section*{Conclusions}
We report the interstellar identification of 1-cyanopyrene, a 4-ring PAH substituted with a \ce{CN} group, in GOTHAM observations of the cold molecular cloud TMC-1. We derive a high abundance of 1-cyanopyrene, and estimate that the unsubstituted pyrene potentially locks up $0.03 - 0.3\,\%$ of carbon in TMC-1. Cyanopyrene is the largest PAH discovered yet in dense molecular clouds and is present in similar abundances to the single-ring aromatic cyanobenzene (benzonitrile), suggesting an unexplored reservoir of larger PAHs in the ISM. This discovery delivers additional support for the PAH hypothesis and further evidence of the ubiquitous presence of PAHs in space.  The ties to solar system measurements of PAHs suggest a substantial inheritance of PAHs produced in the cold (T $\sim$10~K) conditions that occur $\sim$1~Myr before star birth.  This represents a promising source of carbon for forming terrestrial worlds which are supplied carbon in the form of solid state organics \cite{2021SciA....7.3632L}.   These organics are potentially created in its own natal cloud and hints at a ubiquitous source of carbon for forming planetary systems.



\section*{Acknowledgments}  The authors thank T. Lamberts for helpful discussions.  The National Radio Astronomy Observatory is a facility of the National Science Foundation operated under cooperative agreement by Associated Universities, Inc. The Green Bank Observatory is a facility of the National Science Foundation operated under cooperative agreement by Associated Universities, Inc.\\

\textbf{Funding:} G.W., D.A.S., and B.A.M. acknowledge the support of the Arnold and Mabel Beckman Foundation Beckman Young Investigator Award.  M.D., Z.T.P.F., M.S.H., and B.A.M. acknowledge support from the Schmidt Family Futures Foundation. A.N.B. acknowledges the support of NSF Graduate Research Fellowship grant 2141064. C.X. and B.A.M. acknowledge support of National Science Foundation grant AST-2205126. I.R.C. acknowledges support from the University of British Columbia, the Natural Sciences and Engineering Research Council of Canada, the Canada Foundation for Innovation and the B.C. Knowledge Development Fund (BCKDF). P.B.C. and M.C.M. are supported by the National Science Foundation award No. AST-2307137. M.A.C. and S.B.C. are supported by the Goddard Center for Astrobiology and by the NASA Planetary Science Division Internal Scientist Funding Program through the Fundamental Laboratory Research work package (FLaRe). H.G. acknowledges support from the National Science Foundation for participation in this work as part of his independent research and development plan. Any opinions, findings, and conclusions expressed in this material are those of the authors and do not necessarily reflect the views of the National Science Foundation. \\

\textbf{Author Contributions:}
All authors edited and reviewed the manuscript.  In addition, G.W. performed spectroscopic experiments, analyzed observational data, conducted quantum chemical calculations, and wrote the manuscript.  I.R.C. analyzed observational data and wrote the manuscript.  P.B.C. performed spectroscopic experiments.  E.A.B. analyzed observational results.  S.Z. performed the synthesis.  A.M.B. performed observations.  A.J.R. performed observations.  C.X. performed observations.  M.C.M. supervised laboratory experiments.  A.E.W. supervised laboratory experiments.  B.A.M. performed observations, analyzed observational data, wrote the manuscript, and designed the project.\\

\textbf{Competing Interests:} The authors declare no competing interests.\\

\textbf{Data and Materials Availability:} All data from our observing program GOTHAM are now available to the scientific community through the NRAO and GBO archives at  https://data.nrao.edu/portal/ under project codes GBT17A\nobreakdash-164, GBT17A\nobreakdash-434, GBT18A\nobreakdash-333, GBT18B\nobreakdash-007, GBT19B\nobreakdash-047, AGBT20A\nobreakdash-516, AGBT21A\nobreakdash-414, and AGBT21B\nobreakdash-210. Calibrated and reduced observational data windowed around the reported transitions; the full catalog of 1-cyanopyrene, including spectroscopic properties of each transition; and the partition function used in the MCMC analysis are available in a Harvard Dataverse repository OR zenodo.\\

\section*{Supplementary Materials}
Materials and Methods\\
Figs. S1 to S9\\
Tables S1 to S5\\
Movie S1\\
References \textit{(62 -- 73)}

\clearpage

\setcounter{page}{1}

Supplementary Materials for

\begin{center}
\textbf{Discovery of interstellar 1-cyanopyrene: a four-ring \\polycyclic aromatic hydrocarbon in TMC-1}     
\end{center}

Gabi Wenzel, Ilsa R. Cooke, P. Bryan Changala, Edwin A. Bergin, Shuo Zhang, Andrew M. Burkhardt, Alex N. Byrne, Steven B. Charnley, Martin A. Cordiner, Miya Duffy, Zachary T. P. Fried, Harshal Gupta, Martin S. Holdren, Andrew Lipnicky, Ryan A. Loomis,  Hannah Toru Shay, Christopher N. Shingledecker, Mark A. Siebert, D. Archie Stewart, Reace H. J. Willis, Ci Xue, Anthony J. Remijan, Alison E. Wendlandt, Michael C. McCarthy, and Brett A. McGuire

\textbf{This PDF file includes:}


\vspace{-1em}
Materials \& Methods

\vspace{-1em}
Figures S1 -- S9

\vspace{-1em}
Tables S1 -- S5

\vspace{-1em}
Movie S1

\vspace{-1em}
References \textit{(62 -- 73)}

\clearpage
\part*{Materials \& Methods}
\renewcommand{\thefigure}{S\arabic{figure}}
\renewcommand{\thetable}{S\arabic{table}}
\renewcommand{\theequation}{S\arabic{equation}}
\setcounter{figure}{0}
\setcounter{table}{0}
\setcounter{equation}{0}
\setcounter{section}{0}
\section{Synthesis}

\begin{figure}[htb]
    \centering
    \includegraphics[width=0.5\textwidth]{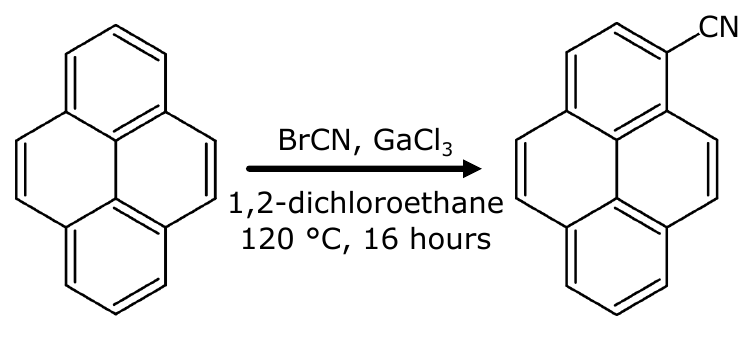}
    \caption{\textbf{Synthesis of 1-cyanopyrene}. Using pyrene as a starting material, 1-cyanopyrene was synthesized in a gallium-catalyzed cyanation reaction.}
    \label{fig:synthesis}
\end{figure}

In a nitrogen-filled glove box, pyrene ($3.0\,\mathrm{g}$, $14.85\,\mathrm{mmol}$) and \ce{GaCl3} ($262\,\mathrm{mg}$, $1.49\,\mathrm{mmol}$) were dissolved in 1,2-dichloroethane ($30\,\mathrm{mL}$). Then, cyanogen bromide (\ce{BrCN}, $3.14\,\mathrm{g}$, $29.69\,\mathrm{mmol}$) was added to the solution to induce a gallium-catalyzed cyanation reaction as reported previously \cite{okamoto2012}. After stirring at $120\,^\circ\mathrm{C}$ for $16\,\mathrm{hours}$, the reaction mixture was quenched with 1M \ce{NaOH} aq. and extracted with \ce{CH2Cl2} ($3\times100\,\mathrm{mL}$). The combined organic layer was dried over \ce{Na2SO4}, filtered, and concentrated under vacuum. The residue was purified by column chromatography on silica gel (hexane/\ce{CH2Cl2} = 2/1) to give the 1-cyanopyrene as a white solid ($1.8\,\mathrm{g}$, 53\,\%). $^1$H NMR ($500\,\mathrm{MHz}$, \ce{CDCl3}) $\delta$ 8.27 (d, $J$ = 9.0 Hz, 1H), 8.24 – 8.18 (m, 2H), 8.16 – 8.09 (m, 3H), 8.08 – 7.98 (m, 2H), 7.94 (d, $J$ = 8.9 Hz, 1H). $^{13}$C NMR (126 MHz, \ce{CDCl3}) $\delta$134.10, 132.85, 130.78, 130.46, 130.45, 130.40, 129.47, 127.06, 126.99, 126.96, 126.82, 124.35, 123.83, 123.83 (overlapping), 123.39, 118.90, 105.51.

\begin{figure}[htb]
   \centering
    \includegraphics[width=0.95\textwidth]{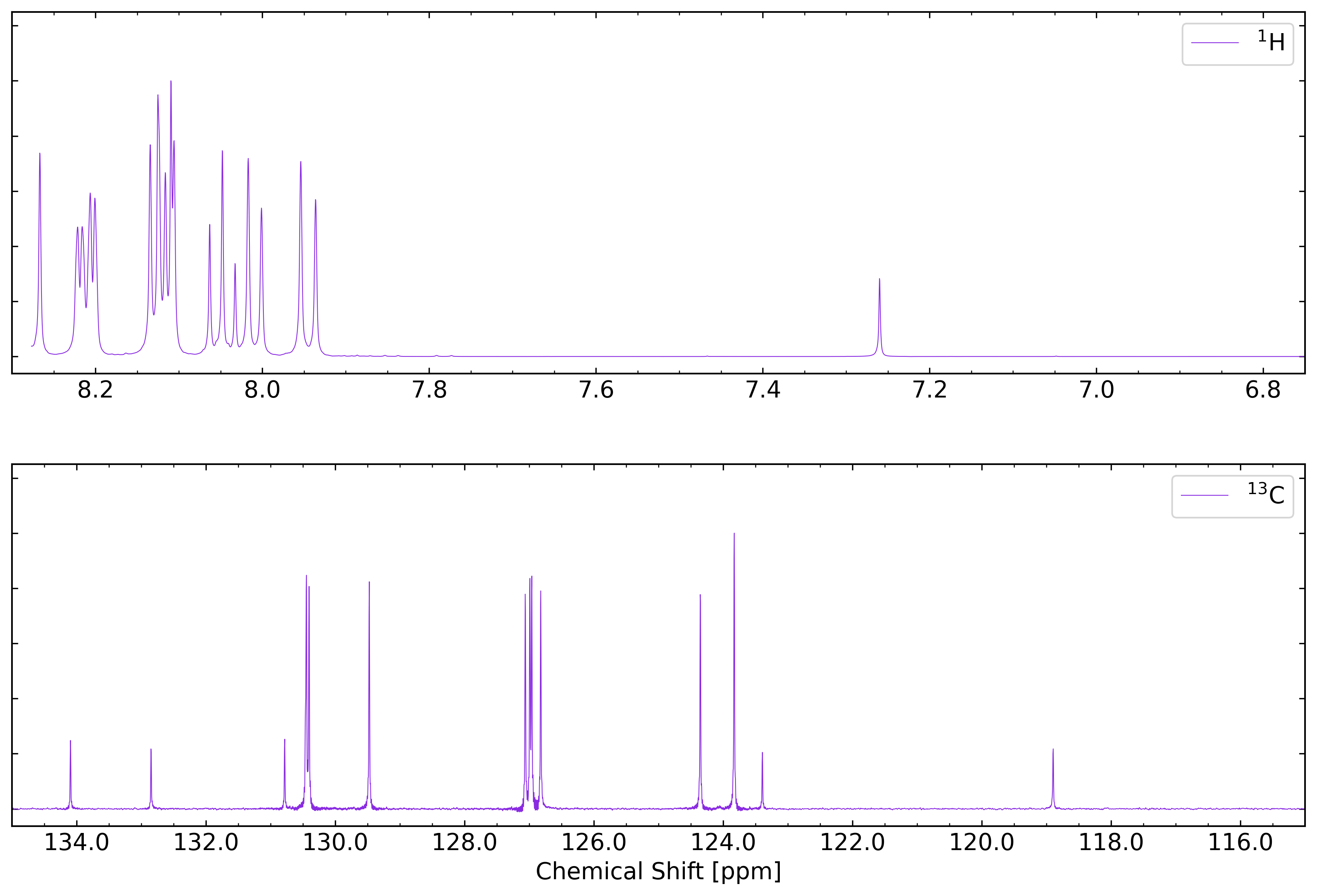}
    \caption{\textbf{Nuclear magnetic resonance (NMR) spectroscopy of synthesized 1-cyanopyrene.} 1-cyanopyrene $^{1}$H NMR (upper panel) and $^{13}$C NMR (lower panel).}
    \label{fig:NMR}
\end{figure}

\section{Rotational Spectroscopy}

1-cyanopyrene was blended with anthracene as a binder material in a 1:1 ratio to produce a homogeneous mixture. Anthracene is rotational spectroscopically invisible due to its lack of a permanent dipole moment. $500\,\mathrm{mg}$ of the mixture were pressed with $3\,\mathrm{tons}$ of force in a hydraulic press into a 0.25''-diameter cylindrical sample rod which was mounted in the laser ablation source, located $8\,\mathrm{mm}$ downstream of a pulsed solenoid valve backed with neon at a pressure of $2.5\,\mathrm{kTorr}$. The rod was ablated with the second harmonic of a Continuum Surelite Nd:YAG laser at $532\,\mathrm{nm}$ with a $50\,\mathrm{mJ}$ pulse energy. The laser pulse was synchronized to fire during the $\leq1\,\mathrm{ms}$ opening time of the solenoid valve. The gas expanded supersonically through the nozzle, carrying the laser-ablated 1-cyanopyrene along the axis of the cavity-enhanced Fourier transform microwave (FTMW) spectrometer \cite{grabow2005, crabtree2016} at a repetition rate of $5\,\mathrm{Hz}$. The expansion cooled the molecules to a rotational temperature of ${\sim}2\,\mathrm{K}$.

Searches for 1-cyanopyrene were performed with the FTMW spectrometer based on its predicted rotational spectrum. Using the open-source quantum chemical package Psi4 \cite{smith2020}, the geometry of 1-cyanopyrene was initially optimized at the B3LYP/6-311++G(d,p) level of theory and then at the M06-2X/6-31+G(d) as recommended in earlier work \cite{lee2020} to determine the rotational constants $A$, $B$, and $C$. The calculated constants are presented in Table \ref{tab:rotconst} and are in good agreement with the constants determined by the so-called `Lego brick' approach applied previously \cite{ye2022}. 

\begin{table}[htb!]
    \centering
    \caption{Spectroscopic constants of 1-cyanopyrene. All dimensional values are reported in MHz.}
    \begin{tabular}{lS[table-format=3.10]S[table-format=3.5]lS[table-format=3.5]cccc}
    \hline
        & \multicolumn{2}{c}{This work} & & \multicolumn{1}{c}{Ye et al. (2022)} \\
        
        {Parameter} & {Experimental} & {M06-2X/6-31+G(d)} & &{`Lego brick' \cite{ye2022}} \\
    \cline{1-3}
    \cline{5-6}
        &  &  &  &  &  \\
        $A$ &  843.140191(128) & 844.3866 & & 843.396  \\
        $B$ &  372.500175(56) & 372.4787 & & 372.076  \\
        $C$ &  258.4249175(164) & 258.4641 & & 258.233  \\
        $\Delta_J \times 10^6$ &  2.240(80) & &  \\
        $\Delta_{JK} \times 10^6$ &  -5.52(101) & &  \\
        $\Delta_K \times 10^6$ & $\cdots$ &  &   \\
        $\delta_J \times 10^6$ & 0.826(40) &  &  \\
        $\delta_K \times 10^6$ & 5.26(74) &  & \\
        $\chi_{aa}$ & -1.8352(228) & -1.8839 & & \\
        $\chi_{bb}$ & -0.0437(163) & -0.0649 & & \\
        $\chi_{ab}$ & 4.01(48) & 1.9488 & & \\
         &  &  &  &  & \\
        $N_\mathrm{lines}$ & 267 &  & \\
        $\sigma_\mathrm{fit} \times 10^3$ & 2.074 &  &  & \\
        $(J, K_\mathrm{a})_\mathrm{max}$ & {$(31, 7)$} &  &  & \\
        \hline
    \end{tabular}
    \label{tab:rotconst}
\end{table}

We used SPCAT/SPFIT in Pickett’s CALPGM suite of programs \cite{pickett1991} to derive the spectroscopic constants of 1-cyanopyrene by least-squares fitting with a Watson Hamiltonian (A-reduced, $I^r$ representation). Quartic centrifugal distortion and $^{14}$N nuclear electric quadrupole hyperfine coupling constants were included in the fitting procedure. Initial starting values for $\chi_{aa}$, $\chi_{bb}$, and $\chi_{ab}$ were determined by comparison of the calculated structures of benzonitrile (cyanobenzene) and 1-cyanopyrene, rotating the former into the principal axis coordinate system of the latter. The resulting constants are listed in Table \ref{tab:rotconst} and are in remarkable agreement with the predicted values. 

\begin{table}[htb!]
    \centering
    \caption{Partition function for 1-cyanopyrene as determined by SPCAT and used in the MCMC analysis.}
    \begin{tabular}{S[table-format=3.3]S[table-format=3.5]S[table-format=3.4]}
    \hline
        {Temperature [K]} & & {Partition function} \\
    \hline
       & \\
1.0  &&   1782.6002 \\
1.5  &&  3270.3246\\
2.0  && 5031.5211\\
2.5  &&  7028.8560\\
3.0  &&   9237.1247\\
3.5  &&  11637.8175\\
4.0  &&  14216.5893\\
4.5  &&  16961.8971\\
5.0  &&   19864.1903\\
5.5  &&  22915.3959\\
6.0  &&  26108.5721\\
6.5 &&   29437.6667\\
7.0  &&  32897.3415\\
7.5  &&  36482.8422\\
8.0  &&  40189.8990\\
8.5  &&  44014.6497\\
9.375 && 50980.7580\\
18.75 && 144163.7503\\
37.5  && 406943.9665\\
75.0  && 1101090.4020\\
150.0 && 2505079.7095\\
225.0 && 3592366.0892\\
275.0 && 5146401.1087\\
300.0 && 4397120.2574\\
400.0 && 5173292.2794\\
500.0 && 5731202.3419\\
        \hline
    \end{tabular}
    \label{tab:partfunc}
\end{table}

\section{Observations and Data Reduction}

\begin{figure}[htb!]
    \centering
    \includegraphics[width=0.95\linewidth]{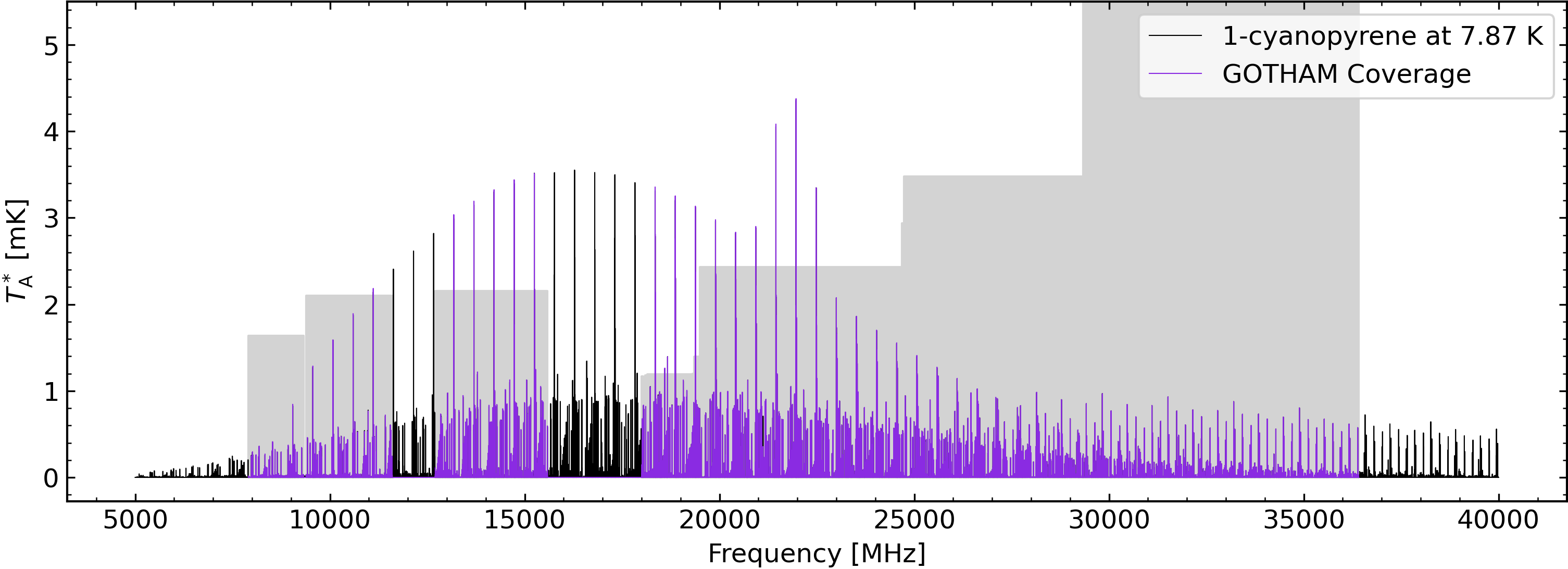}
    \caption{\textbf{Coverage of 1-cyanopyrene lines in the GOTHAM  data set.} Rotational spectrum of 1-cyanopyrene simulated at the derived excitation temperature of $7.87\,\mathrm{K}$ (black). The lines that are covered in the GOTHAM data are depicted in violet. The grey shaded areas indicate the noise level of the observation averaged in each data chunk, highlighting several frequency ranges in which 1-cyanopyrene individual line detection are viable.}
    \label{fig:coverage}
\end{figure}

The GOTHAM project was described in detail previously \cite{mcguire2020,mcguire2021}. Spectra were collected using the VErsatile GBT Astronomical Spectrometer (VEGAS) on the 100\,m  Robert C. Byrd Green Bank Telescope (GBT). In this study, we use the data set presented in detail in refs. \cite{sita2022,cooke2023}, that includes observations until May 2022, and this section is heavily reproduced from these references. Briefly, the data set covers the entirety of the X-, Ku-, K-, and most of the Ka-receiver bands with nearly continuous coverage from $7.9-11.6\,\mathrm{GHz}$, $12.7-15.6\,\mathrm{GHz}$, and $18.0-36.4\,\mathrm{GHz}$, yielding a total bandwidth of $24.9\,\mathrm{GHz}$. Observations were performed with the project codes GBT17A\nobreakdash-164, GBT17A\nobreakdash-434, GBT18A\nobreakdash-333, GBT18B\nobreakdash-007, GBT19B\nobreakdash-047, AGBT20A\nobreakdash-516, AGBT21A\nobreakdash-414, and AGBT21B\nobreakdash-210. Pointing was performed on the cyanopolyyne peak (CP) in TMC-1 at (J2000) $\alpha$~=~04$^h$41$^m$42.50$^s$ $\delta$~=~+25$^{\circ}$41$^{\prime}$26.8$^{\prime\prime}$ and ON/OFF-source spectra were collected using position-switching to an emission-free position offset by $1^{\circ}$. Depending on the receiver and hence frequency range, re-pointing and focusing was carried out every $1-2$\,hours, primarily on the calibrator J0530+1331. The flux was calibrated using an internal noise diode and Karl G. Jansky Very Large Array (VLA) observations of the same calibrator used for pointing. This results in a flux uncertainty of approximately $20\,\%$ \cite{sita2022,cooke2023}. The receivers were set to record data with a uniform frequency resolution of $1.4\,\mathrm{kHz}$, corresponding to $0.05 - 0.01\,\mathrm{km\,s^{-1}}$ in velocity. The coverage of the GOTHAM data is depicted in Fig. \ref{fig:coverage} overlaid with the simulated rotational spectrum of 1-cyanopyrene at $7.87\,\mathrm{K}$. The area shaded in grey represents the averaged root mean square (RMS) noise level, on the order of ${\sim}2-20\,\mathrm{mK}$, in each data chunk, while the overlap between GOTHAM coverage and 1-cyanopyrene lines is shown in violet. The RMS increase toward higher frequencies is due to less total integration time in those frequency ranges.

\section{MCMC Analysis \& Results}

\begin{figure}[htb!]
    \centering
    \includegraphics[width=1\linewidth]{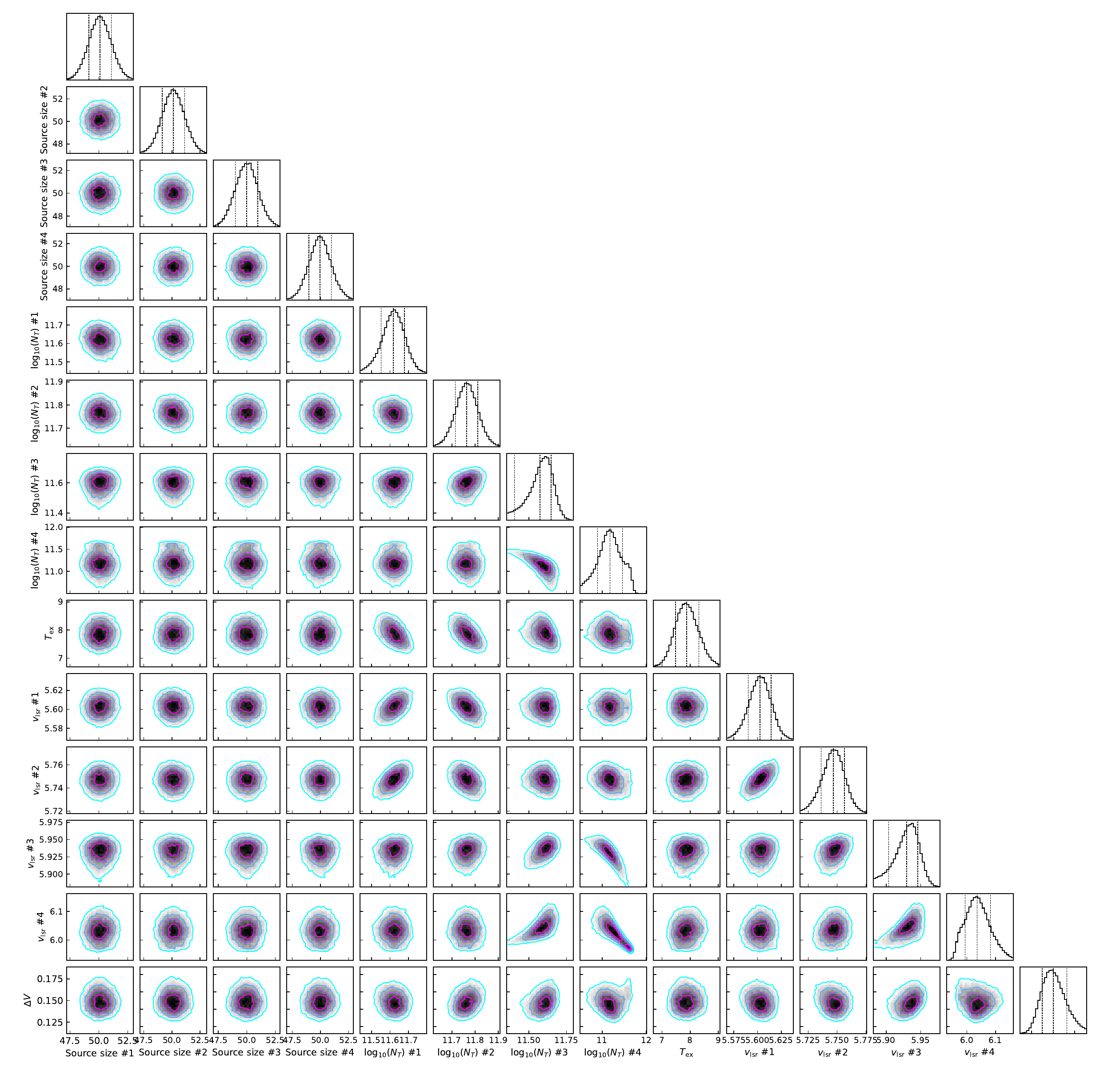}
    \caption{\textbf{Corner plot of MCMC analysis for 1-cyanopyrene.} This shows the parameter covariances on the off-diagonal and marginalized posterior distributions considered in the marginalized posterior distributions on-diagonal. The $16^\mathrm{th}$, $50^\mathrm{th}$, and $84^\mathrm{th}$ confidence intervals (corresponding to $\pm 1\,\sigma$ for a Gaussian posterior distribution) are shown as vertical lines on the diagonal.}
    \label{fig:cornerplot}
\end{figure}

Considering four different Doppler components of TMC-1, each with independent source size, velocity in the local standard of rest, $v_\mathrm{lsr}$, and column densities, $N_T$, together with a uniform excitation temperature, $T_\mathrm{ex}$, and linewidth, $\Delta V$, results in 14 parameters to be fitted for any molecule. MCMC analysis has proven to be a powerful tool to approximate complex probability distributions in many components, see refs. \cite{loomis2021,mcguire2021} for details on the method. Table \ref{tab:MCMCpriors} lists the adopted priors for the 1-cyanopyrene MCMC analysis. Posterior probability distributions for each parameter and their covariances were generated with 100 walkers with 10,000 samples. From this analysis, we extract the source-dependent molecular parameters for 1-cyanopyrene directly from the GOTHAM data as reported in Table \ref{tab:MCMCresults}. It is important to note that these results are to some extent a function of the constraints we apply in setting up the priors, which are in agreement with previously found parameters. The priors for $N_T$ and $T$ are chosen to be a uniform (unconstrained within the minima and maxima) distribution, whereas the remaining parameters have fairly tightly constrained Gaussian distributed priors centered at their known values, determined by prior observations of 
benzonitrile and the cyanonaphthalenes \cite{mcguire2021}. The covariance plots of the 14 parameters resulting from our MCMC analysis are presented in Fig. \ref{fig:cornerplot}. 

\begin{table}[htb!]
    \caption{Priors used for the MCMC analysis, where $N(\mu,\sigma^2)$ denotes a normal (or Gaussian) parameter distribution with mean, $\mu$, and variance, $\sigma^2$, and $U\{a,b\}$ denotes a uniform (unweighted) parameter distribution between $a$ and $b$. The minimum and maximum values for each distribution are noted beneath it. These priors (except source size*) were derived from the marginalized posterior for the MCMC analysis of 2-cyanonaphthalene \cite{mcguire2021}.}
    \centering
    \begin{tabular}{cccccccccc}
      Component &  $v_\mathrm{lsr}$	&	Size*	&	$\mathrm{log_{10}}(N_T)$	&	$T_\mathrm{ex}$	&	$\Delta V$	\\
	No. & [$\mathrm{km\,s^{-1}}$] &[$^{\prime\prime}$]	&	[$\mathrm{cm}^{-2}$]	&	[$\mathrm{K}$]	&	[$\mathrm{km\,s^{-1}}$]\\
	\midrule
	1 & {$N(5.589,0.05)$} & \multirow{4}{*}{$N(50,1)$}	 &  	 \multirow{4}{*}
 {$U\{\mathrm{a,b}\}$} &
 \multirow{4}{*}{$U\{\mathrm{a,b}\}$}	 & 	 \multirow{4}{*}{$N(0.155,0.05)$}\\
	2 &	{$N(5.767,0.05)$} &	 & 	 & 	 & 	 \\
        3 & {$N(5.889,0.05)$} &	 & 	 &   & 	 \\
	4 &	{$N(6.023,0.05)$} &  &	 & 	 & 	 \\
 \midrule
        Min & $0.0$ & $5$ & $10.5$ & $4.0$ & $0.05$ \\
        Max & $10.0$ & $100$ & $13.0$ & $10.0$ & $0.25$ \\
	\midrule
\\
    \end{tabular}\\
    {\small{* We note that due to the GOTHAM survey's lack of spatial information and degeneracy with other fitting parameters, we assumed a narrow source size distribution as listed in Table \ref{tab:MCMCpriors}, and chose not to report errors deduced from the MCMC fitting analysis as reported in Table \ref{tab:MCMCresults} to avoid confusion about the accuracy of this parameter.}\par}
    \label{tab:MCMCpriors}
\end{table}

\begin{table}[htb!]
    \centering
    \caption{Summary statistics of the marginalized posterior for the MCMC analysis of 1-cyanopyrene using the GOTHAM data set together with priors and parameter distribution as reported in Table \ref{tab:MCMCpriors}.}
    \begin{tabular}{cccccccccc}
      Component &  $v_\mathrm{lsr}$	&	Size	&	$N_T$	&	$T_\mathrm{ex}$	&	$\Delta V$	\\
	No. & [$\mathrm{km\,s^{-1}}$] & [$^{\prime\prime}$]	&	[$10^{11}\,\mathrm{cm}^{-2}$]	&	[$\mathrm{K}$]	&	[$\mathrm{km\,s^{-1}}$]\\
    \midrule
	1 &	$5.603^{+0.012}_{-0.012}$	 & 	$50$	 & 	$4.15^{+0.62}_{-0.59}$	 & 	 \multirow{4}{*}{$7.87^{+0.43}_{-0.40}$}	 & 	 \multirow{4}{*}{$0.150^{+0.016}_{-0.013}$}\\
	2 &	$5.747^{+0.010}_{-0.011}$	 & 	$50$	 & 	$5.81^{+0.68}_{-0.61}$	 & 	 & 	 \\		
        3 & $5.930^{+0.016}_{-0.026}$	 & 	$49$	 & 	$3.75^{+0.70}_{-1.21}$	 & 	 & 	 \\
	4 &	$6.036^{+0.046}_{-0.042}$	 & 	$49$	 & 	$1.49^{+1.38}_{-0.71}$	 & 	 & 	 \\
	\midrule
		\multicolumn{6}{c}{$N_T(\mathrm{Total}):\;1.52^{+0.18}_{-0.16}\times 10^{12}\,\mathrm{cm}^{-2}$}\\
    \end{tabular}
    \label{tab:MCMCresults}
\end{table}

\clearpage

\section{Robustness Tests for Stacking and Matched Filtering Analysis}

The velocity-stacking and matched filtering analysis is described in the main text and in more detail in refs. \cite{loomis2021,mcguire2021}.  Care was taken not to include any possible interlopers, discarding every spectral window with emission features above an SNR of $4\,\sigma$.  In the following sections, we provide a video showcasing the stacking and matched filtering analysis on a \textit{line by line} basis. Further, we describe the robustness tests we performed on the technique in order to confirm our 1-cyanopyrene detection of $14.3\,\sigma$.

\subsection{Number of Lines in Stack}

\begin{figure}[htb!]
    \centering
    \includegraphics[width=1\linewidth]{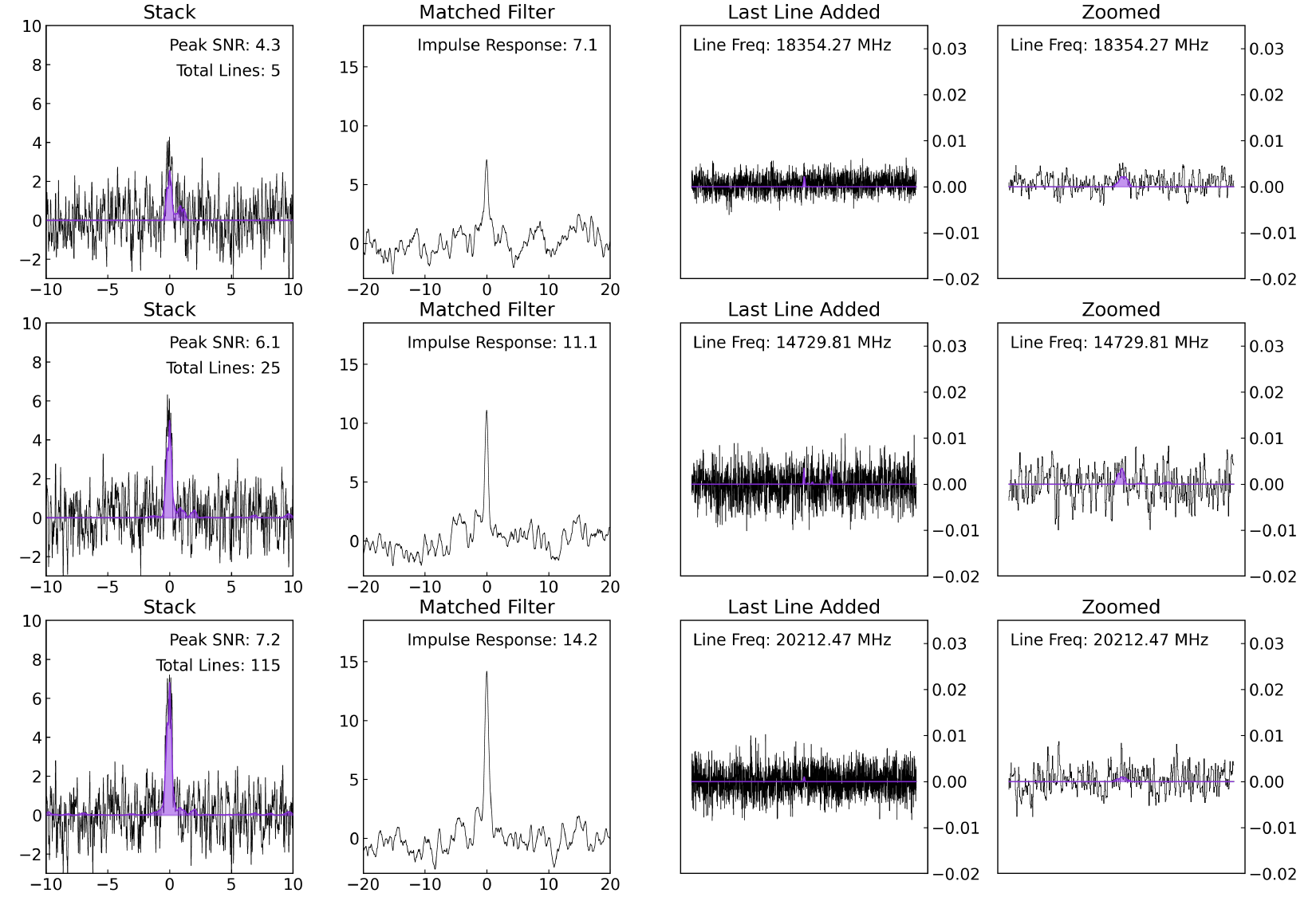}
    \caption{\textbf{Snapshots of the video demonstration of the stacking and matched filtering analysis.} Snapshots show the stack and fitted simulated spectra (left panel, black and violet respectively) and corresponding matched filter response (second left panel) together with the spectral window chosen around each considered line that makes it into the stack (both panels on the right). Snapshots were taken at a total number of lines in the stack of 5 (top row), 25 (middle row), and 145 (bottom row).}
    \label{fig:video}
\end{figure}

\begin{figure}[htb!]
    \centering
    \includegraphics[width=0.75\linewidth]{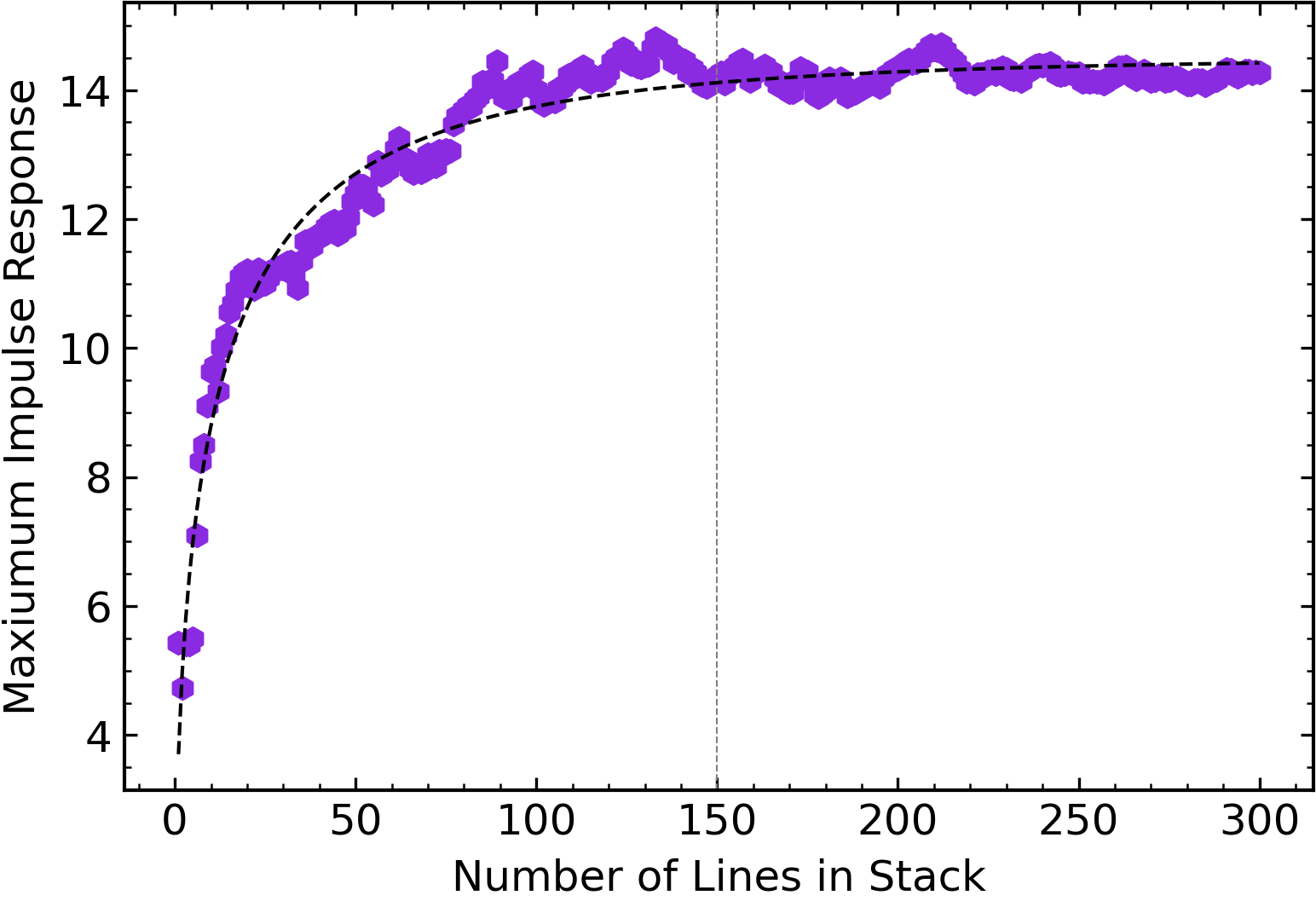}
    \caption{\textbf{Determining the number of lines to be used in the stack.} The maximum matched filter response is shown as a function of the number of brightest SNR lines included in the velocity stack. The black dashed line represents the $E(N) = A(1-e^{-k\sqrt{N}})$ trend (see text).}
    \label{fig:mf_evolution}
\end{figure}

We compiled a video consisting of snapshots on a \textit{line by line} basis of the velocity-stacking and matched filtering procedure. Example shots from that video are presented in Fig.~\ref{fig:video}, demonstrating the spectral window selected centered around each line considered (panels on the right), the behavior of the velocity-stacks of the GOTHAM data (left panel, black) and the 1-cyanopyrene simulated spectrum (left panel, violet), together with the matched filter response (second left panel). The evolution, $E$, of the matched filter response with the number of highest SNR lines, $N$, included in the stack is depicted in fig. \ref{fig:mf_evolution} and follows approximately a $E(N) = A(1-e^{-k\sqrt{N}})$ trend, where $\lim_{N\to\infty} E(N) = A$ is the maximum filter response to be reached with rate $k$. Together, these demonstrate that the significant detections found by the matched filter response and velocity stacks originate from the amassing of many of the highest SNR lines and do not include spurious features.  

\subsection{Jack-knife Test}

\begin{figure}[htb]
    \centering
    \includegraphics[width=0.95\linewidth]{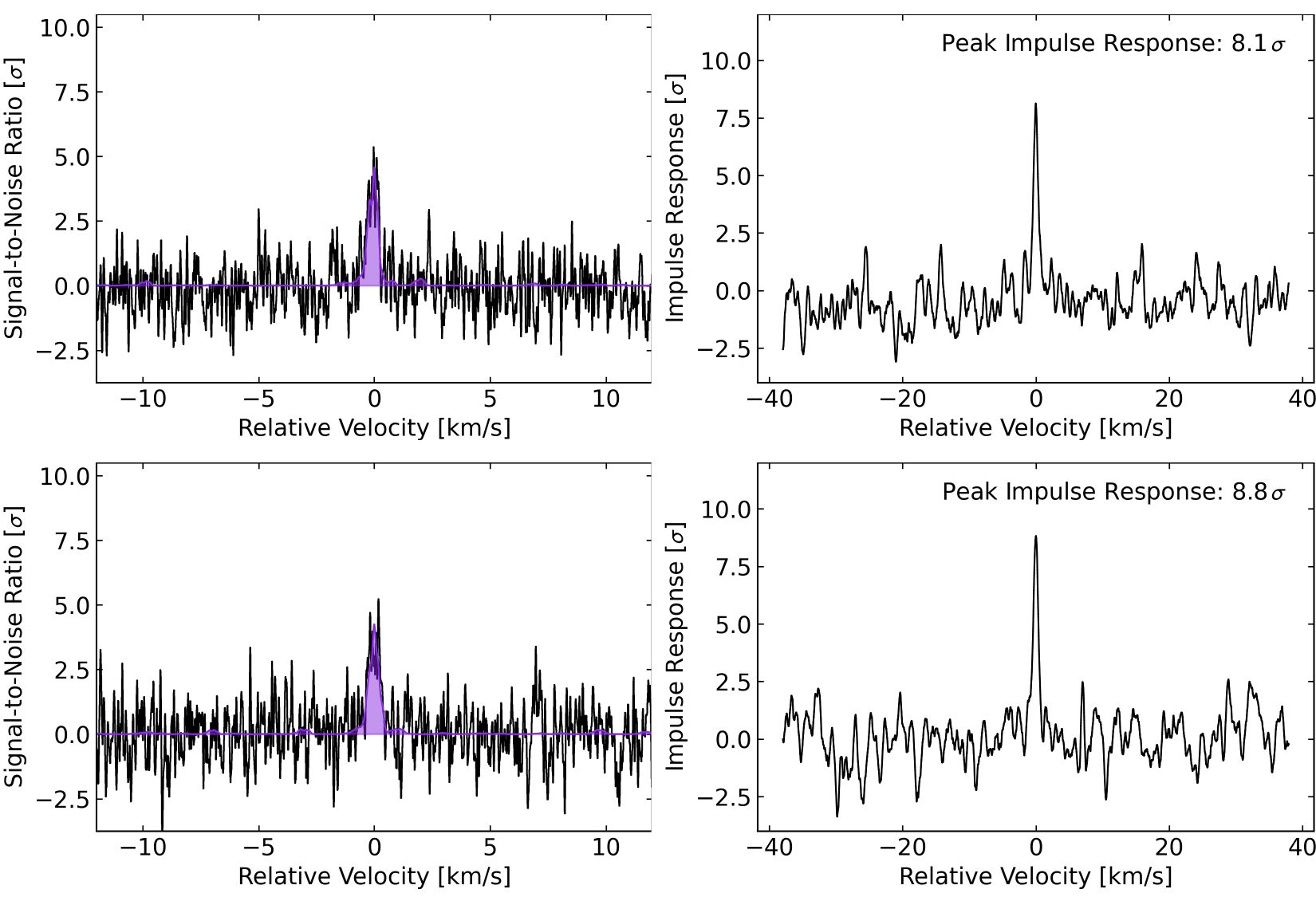}
    \caption{\textbf{Results of the jack-knife test for 1-cyanopyrene.} Upper panels show the stack and matched filtering analysis performed on the first half (even number of lines) while lower panels depict the second half (odd number of lines) of the total number of 150 stacked lines.}
    \label{fig:jackknife}
\end{figure}

In order to test the robustness of our velocity-stacking and matched filter analysis of the 1-cyanopyrene signal in the GOTHAM data, we performed a jack-knife test. With a line cutoff set by fig. \ref{fig:mf_evolution}, the 150 brightest SNR lines in the simulated 1-cyanopyrene spectrum were stacked against the respective observational data, creating the ``full'' stack resulting in a matched filter response of $14.3\,\sigma$ as shown in Fig. \ref{fig:stack+mf}. We now divide the 150 lines into two sets, considering every other line (each observable line as seen in fig. \ref{fig:coverage} that might contain multiple hyperfine components or different $K$-components) and so creating an ``even'' and an ``odd'' dataset of lines in the spectrum. Stacking these datasets separately against the GOTHAM data delivers the velocity-stacks and matched filters as depicted in fig. \ref{fig:jackknife}, yielding an impulse response of $8.1\,\sigma$ and $8.8\,\sigma$, respectively. If we expect the signal to be real and to originate only from the molecule considered, the matched filter responses of the ``half'' stacks should be approximately $2^{-\frac{1}{2}}\times$ of the matched filter response of the ``full'' stack. Adding the responses of both ``half'' stacks in quadrature yields $12.0\,\sigma$, comparable with the value from the ``full'' stack. 

\subsection{Accuracy of Rotational Constants}

\begin{figure}[htb!]
    \centering
    \includegraphics[width=0.75\linewidth]{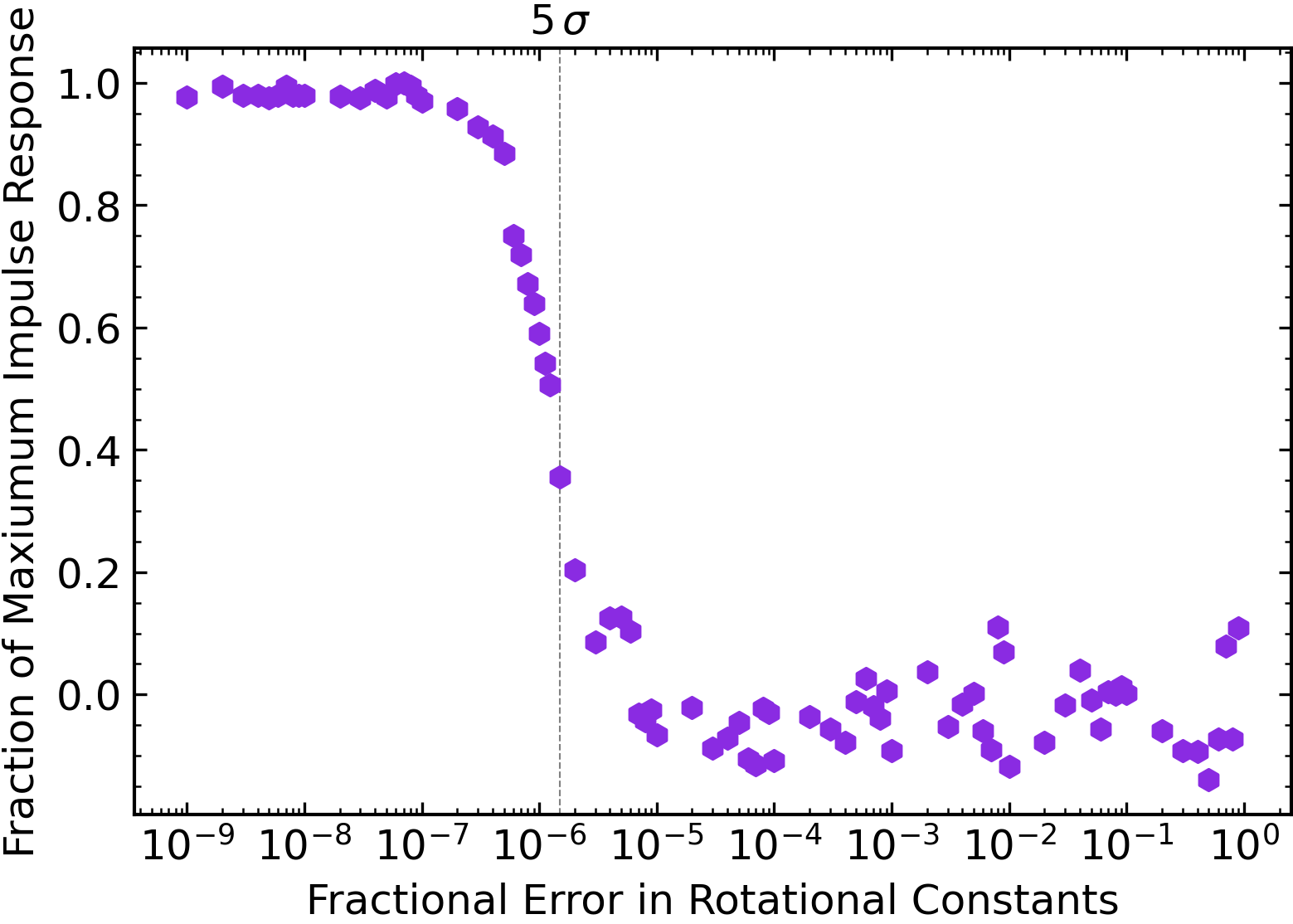}
    \caption{\textbf{Accuracy test of rotational constants.} Normalized matched filter response as a function of fractional error applied to the rotational constants $A$, $B$, and $C$ for 1-cyanopyrene. The dashed line denotes when the matched filter response becomes significant ($>\!\!5\,\sigma$) at an accuracy of $1.5\,\mathrm{ppm}$.}
    \label{fig:accuracy}
\end{figure}

The search for 1-cyanopyrene in TMC-1 demonstrated yet again how crucial accurate spectral catalogs including, if applicable, hyperfine splitting are. To test for potential false positive matched filter responses for 1-cyanopyrene, we introduced a fractional error of varying magnitude into our experimentally obtained rotational constants $A$, $B$, and $C$. We then perform the velocity-stacking and matched filtering analysis using these flawed rotational constants together with the same parameters derived from the MCMC model fitting (see Table \ref{tab:MCMCresults}) of the true catalog (see Table \ref{tab:rotconst}). The results are presented in fig. \ref{fig:accuracy}. We find that the impulse response only becomes significant ($>\!\!5\,\sigma$) at an accuracy of $1.5\,\mathrm{ppm}$, while most of the signal is recovered at $100\,\mathrm{ppb}$ and all is recovered at $10\,\mathrm{ppb}$. This is consistent with previous robustness tests performed on benzonitrile and the cyanonaphthalene isomers \cite{mcguire2021}. We thus conclude that the detected molecule is indeed 1-cyanopyrene.

\subsection{Carbon Budget}\label{SIbudget}

Astrochemical models can be used to predict the H:CN ratio, but they rely on accurate rate coefficients and product-branching ratios under TMC-1 conditions, most of which are unavailable. Since most PAHs lack permanent dipole moments, it is challenging to obtain observational constraints on this ratio; however, the cyanoindene/indene pair, as well as the 1-cyanocyclopentadiene/cyclopentadiene and 2-cyanocyclopentadiene/cyclopentadiene pairs, have all been detected in TMC-1, and we adopt their ratios as rough boundaries. The observed H:CN ratios were 14.5:1 (\ce{C5H6}:1-\ce{C5H5CN}), 43:1 (\ce{C9H8}:2-\ce{C9H7CN}) and 63:1 (\ce{C5H6}:2-\ce{C5H5CN}). Due to the large uncertainty on the H:CN ratio, we assume a lower limit of 10:1 and an upper limit of 100:1. Analysis of the bulk H:CN ratio in PAHs assumed to be contributing to the UIR bands also indicates a value of $<$1:100, though this may not reflect the fraction nor the size of PAHs in dense clouds where the radiation field strength is much lower\cite{tielens2008}. In addition, the H:CN variation among isomers for cyclopentadiene implies that this ratio may also strongly diffe\-r for the other cyanopyrene isomers. Assuming $N(\mathrm{H_2}) \sim 10^{22}\,\mathrm{cm^{-2}}$, the abundance of 1-cyanopyrene with respect to hydrogen is ${\sim}1.5 \times 10^{-10}$. Applying the estimated H:CN ratio ranging from 10:1 to 100:1, we derive an estimated abundance of pyrene of ${\sim}0.15$--$1.5\,\times 10^{-8}$.

Table \ref{tab:budget} shows the fraction of carbon locked up in select molecules observed in TMC-1, including other aromatic compounds detected in the GOTHAM and QUIJOTE (\emph{Q-band Ultrasensitive Inspection Journey to the Obscure TMC-1 Environment} \cite{cernicharo2021b}) surveys, for comparison to that locked up in 1-cyanopyrene.

The column densities of the 1-, 2-, and 4-ring CN-substituted aromatics do not vary significantly with the number of carbon atoms, all lying within an order-of-magnitude of one another (${\sim}0.2-2\times 10^{12} \mathrm{cm^{-2}}$). One explanation for this trend is the possibility of top-down formation from an extended reservoir of larger PAHs in the ISM. The abundance of carbon locked up in 1-cyanopyrene is larger than for any other CN-substituted aromatic, including benzonitrile. Since cyanopyrene has three distinct isomers, the total cyanopyrene column density may yet be even larger \cite{sita2022}.

\begin{figure}[!htb]
    \centering
    \includegraphics[width=0.8\linewidth]{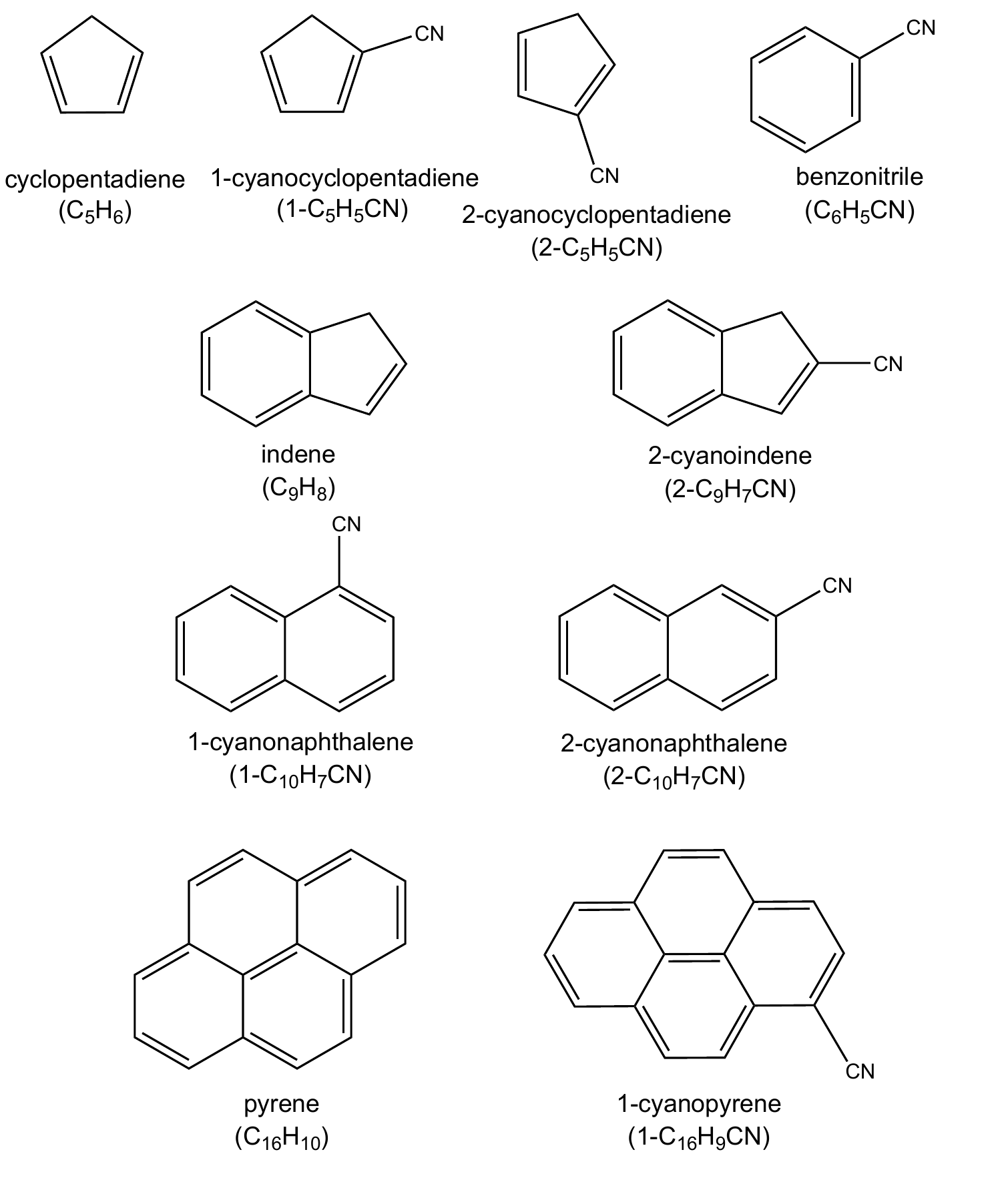}
    \caption{\textbf{Chemical structures, common names and condensed formulae for the molecules in Table S4.}}
    \label{fig:molecularstructures}
\end{figure}

\begin{table}[!htb]
    \centering
    \caption{Column densities of select carbon-bearing molecules observed in TMC-1. The abundance of carbon in each molecule ($f_\mathrm{C}$) in ppm is found by multiplying the abundance, $n$(X)/$n$(H), with the number of carbons in the molecule ($N_\mathrm{C}$). Unfunctionalized hydrocarbons are shown in bold. }    
    \begin{tabular}{c c c c c}
    \toprule
    Molecule      & $N_\mathrm{C}$ &     Column Density     &     {$f_\mathrm{C}$} &  Reference                   \\
                 &   &    [$10^{12}\,\mathrm{cm}^{-2}$]  & [ppm]  &                             \\
     \midrule
    \textbf{\ce{C5H6}}       & 5 & 12$^{+3}_{-3}$  &   6 $\times$ 10$^{-3}$  & \cite{cernicharo2021} \\
    1-\ce{C5H5CN}   & 6 & 0.827$^{+0.09}_{-0.10}$   &  5 $\times$ 10$^{-4}$  & \cite{lee2021}         \\
    2-\ce{C5H5CN}   & 6 & 0.189$^{+0.018}_{-0.015}$ &   1 $\times$ 10$^{-4}$  & \cite{lee2021}         \\
     \ce{C6H5CN}    & 7 &  1.73$^{+0.85}_{-0.10}$  &  1 $\times$ 10$^{-3}$ &  \cite{mcguire2021}\\
    \textbf{\ce{C9H8}}       & 9 &  9.04$^{+0.96}_{-0.96}$    &  8 $\times$ 10$^{-3}$  &  \cite{sita2022}                \\
    2-\ce{C9H7CN}   & 10 &   0.210$^{+0.060}_{-0.046}$ & 2 $\times$ 10$^{-4}$       & \cite{sita2022}           \\
    1-\ce{C10H7CN} & 11 &  0.735$^{+0.330}_{-0.463}$ & 8 $\times$ 10$^{-4}$  &  \cite{mcguire2021}\\
    2-\ce{C10H7CN} & 11 &  0.705$^{+0.450}_{-0.323}$  & 8 $\times$ 10$^{-4}$  &\cite{mcguire2021} \\
    1-\ce{C16H9CN} & 17 & 1.52$^{+0.18}_{-0.16}$ & 3 $\times$ 10$^{-3}$ & This work \\
    \textbf{\ce{C16H10}}* & 16 & 15--150 & 2 ($\times$ 10$^{-2}$ -- 10$^{-1}$) & This work \\    
    \midrule
    \multicolumn{5}{c}{* estimated using the 1-cyanopyrene abundance and taking a H:CN ratio of 10:1 -- 100:1.}\\
    \end{tabular}
    \label{tab:budget}
\end{table}

Our observations probe the gas-phase abundance of cyanopyrene, but it is likely that some cyanopyrene (as well as other PAHs) is present in the solid phase on dust grains due to the low gas and dust temperature in TMC-1. To estimate the abundance of solid-state pyrene, we consider its depletion time ($t_{\textrm{dep}}$) under TMC-1 conditions, 

\begin{equation}\label{eqn:tdep}
    t_{\textrm{dep}} = \frac{1}{\xi \,n_d \, \sigma_d \bar{v}_{g}}
\end{equation}
where $\xi$ is the sticking coefficient (assumed to be unity), $n_d$ and $\sigma_d$ are the dust grain number density and geometric cross section, respectively, and $\bar{v}_{g}$ is the mean thermal speed of pyrene at 10 K. We take $n_d \, \sigma_d \approx 2.1 \times 10^{-21} n_H\, cm^{-1}$ \cite{hollenbach2008}, where $n_H$ is the hydrogen nuclei density in cm$^{-3}$.

Using $n_H = 2.5\times 10^4\,\mathrm{cm^{-3}}$ \cite{pratap1997}, we find a depletion time of ${\sim}2\times 10^5\,\mathrm{yr}$.
If we assume pyrene is produced in the gas phase of TMC-1 and the production has been operative for ${\sim}1\,\mathrm{Myr}$ \cite{navarro-almaida2021,luhman2023}, i.e., 5 depletion times, we estimate that ${\sim}1.2\,\mathrm{ppm}$ is locked up on dust grains (assuming a H:CN of 100). We assume that once pyrene is frozen onto the surface of dust grains, it cannot return to the gas phase due to a lack of efficient non-thermal desorption processes within cold, dark, dense clouds. The total amount of pyrene in the gas and solid phase combined is therefore ${\sim}1.5\,\mathrm{ppm}$, relative to H.

To compare with comets, we determine the abundance of carbon locked up in pyrene relative to Si. Assuming a present-day cosmic abundance for Si of $3.1 \times 10^{-5}$ ($31\,\mathrm{ppm}$) relative to H \cite{nieva2012}, we can infer a C/Si abundance of 0.05 for pyrene. The observed C/Si for comets 67P/Churyumov-Gerasimenko and Halley are 5.4 and 5.7, respectively. Thus, pyrene may account for up to ${\sim}1\,\%$ of the carbon content seen in comets, which further supports that pyrene is an ``island of stability'' among PAHs. 

Lastly, it is expected that a significant fraction of PAHs will be ionized --- with anions hypothesized to be up to $20\,\%$ of the total PAH budget in the diffuse medium \cite{tielens2008} --- and thus, a sizable fraction of carbon may be locked up in charged carbonaceous species. 


\begin{thebibliography}{10}

\bibitem{tielens2008}
A.~Tielens, {\it Annual Review of Astronomy and Astrophysics\/} {\bf 46}, 289
  (2008).

\bibitem{allamandola1985}
L.~J. Allamandola, A.~G. G.~M. Tielens, J.~R. Barker, {\it The Astrophysical
  Journal\/} {\bf 290}, L25 (1985).

\bibitem{leger1984}
A.~L{\'e}ger, J.~L. Puget, {\it Astronomy \& Astrophysics\/} {\bf 500}, 279
  (1984).

\bibitem{allamandola1989}
L.~J. Allamandola, A.~G. G.~M. Tielens, J.~R. Barker, {\it The Astrophysical
  Journal Supplement Series\/} {\bf 71}, 733 (1989).

\bibitem{chabot2020}
M.~Chabot, K.~B{\'e}roff, E.~Dartois, T.~Pino, M.~Godard, {\it The
  Astrophysical Journal\/} {\bf 888}, 17 (2020).

\bibitem{peeters2011}
E.~Peeters, {\it European Astronomical Society Publications Series\/} {\bf 46},
  13 (2011).

\bibitem{chown2024}
R.~Chown, {\it et~al.\/}, {\it Astronomy \& Astrophysics\/} {\bf 685}, A75
  (2024).

\bibitem{dwek1997}
E.~Dwek, {\it et~al.\/}, {\it The Astrophysical Journal\/} {\bf 475}, 565
  (1997).

\bibitem{habart2004a}
E.~Habart, A.~Natta, E.~Kr{\"u}gel, {\it Astronomy \& Astrophysics\/} {\bf
  427}, 179 (2004).

\bibitem{vanderzwet1985}
G.~P. {van der Zwet}, L.~J. Allamandola, {\it Astronomy and Astrophysics\/}
  {\bf 146}, 76 (1985).

\bibitem{leger1985}
A.~L{\'e}ger, L.~D'Hendecourt, {\it Astronomy and Astrophysics\/} {\bf 146}, 81
  (1985).

\bibitem{campbell2015}
E.~K. Campbell, M.~Holz, D.~Gerlich, J.~P. Maier, {\it Nature\/} {\bf 523}, 322
  (2015).

\bibitem{cordiner2019}
M.~A. Cordiner, {\it et~al.\/}, {\it The Astrophysical Journal\/} {\bf 875},
  L28 (2019).

\bibitem{salama1992}
F.~Salama, L.~J. Allamandola, {\it Nature\/} {\bf 358}, 42 (1992).

\bibitem{studier1965}
M.~H. Studier, R.~Hayatsu, E.~Anders, {\it Science\/} {\bf 149}, 1455 (1965).

\bibitem{clemett2010}
S.~J. Clemett, S.~A. Sandford, K.~{Nakamura-Messenger}, F.~H{\"o}rz, D.~S.
  McKAY, {\it Meteoritics \& Planetary Science\/} {\bf 45}, 701 (2010).

\bibitem{zeichner2023}
S.~S. Zeichner, {\it et~al.\/}, {\it Science\/} {\bf 382}, 1411 (2023).

\bibitem{naraoka2000}
H.~Naraoka, A.~Shimoyama, K.~Harada, {\it Earth and Planetary Science
  Letters\/} {\bf 184}, 1 (2000).

\bibitem{zhao2018}
L.~Zhao, {\it et~al.\/}, {\it Nature Astronomy\/} {\bf 2}, 413 (2018).

\bibitem{micelotta2011}
E.~R. Micelotta, A.~P. Jones, A.~G. G.~M. Tielens, {\it Astronomy \&
  Astrophysics\/} {\bf 526}, A52 (2011).

\bibitem{mcguire2021}
B.~A. McGuire, {\it et~al.\/}, {\it Science\/} {\bf 371}, 1265 (2021).

\bibitem{burkhardt2021}
A.~M. Burkhardt, {\it et~al.\/}, {\it The Astrophysical Journal Letters\/} {\bf
  913}, L18 (2021).

\bibitem{cernicharo2021}
J.~Cernicharo, {\it et~al.\/}, {\it Astronomy and astrophysics\/} {\bf 649},
  L15 (2021).

\bibitem{sita2022}
M.~L. Sita, {\it et~al.\/}, {\it The Astrophysical Journal Letters\/} {\bf
  938}, L12 (2022).

\bibitem{cami2010}
J.~Cami, J.~{Bernard-Salas}, E.~Peeters, S.~E. Malek, {\it Science\/} {\bf
  329}, 1180 (2010).

\bibitem{sellgren2010}
K.~Sellgren, {\it et~al.\/}, {\it {The Astrophysical Journal Letters}\/} {\bf
  722}, L54 (2010).

\bibitem{frenklach2024}
M.~Frenklach, A.~W. Jasper, A.~M. Mebel, {\it Physical Chemistry Chemical
  Physics\/}  (2024).

\bibitem{sabbah2017}
H.~Sabbah, {\it et~al.\/}, {\it The Astrophysical Journal\/} {\bf 843}, 34
  (2017).

\bibitem{lecasble2022}
M.~Lecasble, L.~Remusat, J.-C. Viennet, B.~Laurent, S.~Bernard, {\it Geochimica
  et Cosmochimica Acta\/} {\bf 335}, 243 (2022).

\bibitem{aponte2023}
J.~C. Aponte, {\it et~al.\/}, {\it Earth, Planets and Space\/} {\bf 75}, 28
  (2023).

\bibitem{byrne2023}
A.~N. Byrne, C.~Xue, I.~R. Cooke, M.~C. McCarthy, B.~A. McGuire, {\it The
  Astrophysical Journal\/} {\bf 957}, 88 (2023).

\bibitem{mcguire2020}
B.~A. McGuire, {\it et~al.\/}, {\it The Astrophysical Journal Letters\/} {\bf
  900}, L10 (2020).

\bibitem{mcguire2022}
B.~A. McGuire, {\it The Astrophysical Journal Supplement Series\/} {\bf 259},
  30 (2022).

\bibitem{balucani2000}
N.~Balucani, {\it et~al.\/}, {\it The Astrophysical Journal\/} {\bf 545}, 892
  (2000).

\bibitem{cooke2020}
I.~R. Cooke, D.~Gupta, J.~P. Messinger, I.~R. Sims, {\it The Astrophysical
  Journal Letters\/} {\bf 891}, L41 (2020).

\bibitem{leger1987}
A.~L{\'e}ger, L.~{d'Hendecourt}, {\it Polycyclic {{Aromatic Hydrocarbons}} and
  {{Astrophysics}}\/}, A.~L{\'e}ger, L.~{d'Hendecourt}, N.~Boccara, eds.
  (Springer Netherlands, Dordrecht, 1987), pp. 223--254.

\bibitem{naraoka2023}
H.~Naraoka, {\it et~al.\/}, {\it Science\/} {\bf 379}, eabn9033 (2023).

\bibitem{jusko2018a}
P.~Jusko, {\it et~al.\/}, {\it Chemical Physics Letters\/} {\bf 698}, 206
  (2018).

\bibitem{marin2020}
L.~G. Marin, {\it et~al.\/}, {\it The Astrophysical Journal\/} {\bf 889}, 101
  (2020).

\bibitem{campisi2020}
D.~Campisi, A.~Candian, {\it Physical Chemistry Chemical Physics\/} {\bf 22},
  6738 (2020).

\bibitem{jlee2021}
J.~W.~L. Lee, {\it et~al.\/}, {\it Nature Communications\/} {\bf 12}, 6107
  (2021).

\bibitem{wenzel2022}
G.~Wenzel, {\it et~al.\/}, {\it Journal of Molecular Spectroscopy\/} {\bf 385},
  111620 (2022).

\bibitem{garkusha2011}
I.~Garkusha, J.~Fulara, P.~J. Sarre, J.~P. Maier, {\it The Journal of Physical
  Chemistry A\/} {\bf 115}, 10972 (2011).

\bibitem{ye2022}
H.~Ye, S.~Alessandrini, M.~Melosso, C.~Puzzarini, {\it Physical Chemistry
  Chemical Physics\/} {\bf 24}, 23254 (2022).

\bibitem{cooke2023}
I.~R. Cooke, {\it et~al.\/}, {\it The Astrophysical Journal\/} {\bf 948}, 133
  (2023).

\bibitem{Science:Materials}
{See supplementary materials.}

\bibitem{okamoto2012}
K.~Okamoto, M.~Watanabe, M.~Murai, R.~Hatano, K.~Ohe, {\it Chemical
  Communications\/} {\bf 48}, 3127 (2012).

\bibitem{loomis2021}
R.~A. Loomis, {\it et~al.\/}, {\it Nature Astronomy\/} {\bf 5}, 188 (2021).

\bibitem{gratier2016}
P.~Gratier, {\it et~al.\/}, {\it The Astrophysical Journal Supplement Series\/}
  {\bf 225}, 25 (2016).

\bibitem{molsim}
K.~L.~K. Lee, R.~A. Loomis, C.~Xue, S.~{El-Abd}, B.~A. McGuire, Molsim v0.4.0
  (2023).

\bibitem{fuente2019}
A.~Fuente, {\it et~al.\/}, {\it Astronomy \& Astrophysics\/} {\bf 624}, A105
  (2019).

\bibitem{bergin2015}
E.~A. Bergin, G.~A. Blake, F.~Ciesla, M.~M. Hirschmann, J.~Li, {\it Proceedings
  of the National Academy of Sciences\/} {\bf 112}, 8965 (2015).

\bibitem{cernicharo2001}
J.~Cernicharo, {\it et~al.\/}, {\it The Astrophysical Journal\/} {\bf 546},
  L123 (2001).

\bibitem{montillaud2013}
J.~Montillaud, C.~Joblin, D.~Toublanc, {\it Astronomy \& Astrophysics\/} {\bf
  552}, A15 (2013).

\bibitem{arulanantham2024}
N.~Arulanantham, {\it et~al.\/}, {\it The Astrophysical Journal Letters\/} {\bf
  965}, L13 (2024).

\bibitem{andrews2015}
H.~Andrews, {\it et~al.\/}, {\it The Astrophysical Journal\/} {\bf 807}, 99
  (2015).

\bibitem{kaiser2021}
R.~I. Kaiser, N.~Hansen, {\it The Journal of Physical Chemistry A\/} {\bf 125},
  3826 (2021).

\bibitem{parker2012}
D.~S.~N. Parker, {\it et~al.\/}, {\it Proceedings of the National Academy of
  Sciences\/} {\bf 109}, 53 (2012).

\bibitem{zhao2018a}
L.~Zhao, {\it et~al.\/}, {\it The Journal of Physical Chemistry Letters\/} {\bf
  9}, 2620 (2018).

\bibitem{stockett2023}
M.~H. Stockett, {\it et~al.\/}, {\it Nature Communications\/} {\bf 14} (2023).

\bibitem{2021SciA....7.3632L}
J.~{Li}, E.~A. {Bergin}, G.~A. {Blake}, F.~J. {Ciesla}, M.~M. {Hirschmann},
  {\it Science Advances\/} {\bf 7}, eabd3632 (2021).

\bibitem{grabow2005}
J.-U. Grabow, E.~S. Palmer, M.~C. McCarthy, P.~Thaddeus, {\it Review of
  Scientific Instruments\/} {\bf 76}, 093106 (2005).

\bibitem{crabtree2016}
K.~N. Crabtree, {\it et~al.\/}, {\it The Journal of Chemical Physics\/} {\bf
  144}, 124201 (2016).

\bibitem{smith2020}
D.~G.~A. Smith, {\it et~al.\/}, {\it The Journal of Chemical Physics\/} {\bf
  152}, 184108 (2020).

\bibitem{lee2020}
K.~L.~K. Lee, M.~McCarthy, {\it The Journal of Physical Chemistry A\/} {\bf
  124}, 898 (2020).

\bibitem{pickett1991}
H.~M. Pickett, {\it Journal of Molecular Spectroscopy\/} {\bf 148}, 371 (1991).

\bibitem{cernicharo2021b}
J.~Cernicharo, {\it et~al.\/}, {\it Astronomy \& Astrophysics\/} {\bf 652}, L9
  (2021).

\bibitem{lee2021}
K.~L.~K. Lee, {\it et~al.\/}, {\it The Astrophysical Journal Letters\/} {\bf
  910}, L2 (2021).

\bibitem{hollenbach2008}
D.~Hollenbach, M.~J. Kaufman, E.~A. Bergin, G.~J. Melnick, {\it The
  Astrophysical Journal\/} {\bf 690}, 1497 (2008).

\bibitem{pratap1997}
P.~Pratap, {\it et~al.\/}, {\it The Astrophysical Journal\/} {\bf 486}, 862
  (1997).

\bibitem{navarro-almaida2021}
D.~{Navarro-Almaida}, {\it et~al.\/}, {\it Astronomy \& Astrophysics\/} {\bf
  653}, A15 (2021).

\bibitem{luhman2023}
K.~L. Luhman, {\it The Astronomical Journal\/} {\bf 165}, 37 (2023).

\bibitem{nieva2012}
M.-F. Nieva, N.~Przybilla, {\it Astronomy \& Astrophysics\/} {\bf 539}, A143
  (2012).

\end{thebibliography}
\end{document}